\documentclass[aps,prd,nofootinbib,onecolumn,superscriptaddress,preprintnumbers,letterpaper]{revtex4}

\usepackage{amssymb}
\usepackage{amsmath}
\usepackage{epsfig}
\usepackage{hyperref}
\usepackage{comment}
\usepackage{color}
\usepackage{cleveref}
\usepackage{multirow}
\usepackage{capt-of}
\usepackage{siunitx}
\usepackage{graphicx}
\usepackage{gensymb}
\usepackage[caption=false]{subfig}
\usepackage{slashed}
\usepackage{tikz-feynman}
\usepackage{tabularx}

\makeatletter
\def\simgt{\mathrel{\lower2.5pt\vbox{\lineskip=0pt\baselineskip=0pt
           \hbox{$>$}\hbox{$\sim$}}}}
\def\simlt{\mathrel{\lower2.5pt\vbox{\lineskip=0pt\baselineskip=0pt
           \hbox{$<$}\hbox{$\sim$}}}}
\makeatother

\begin{document}
\title{TASI Lectures on Indirect Detection of Dark Matter}
\author{Tracy R. Slatyer}
\affiliation{Center for Theoretical Physics, Massachusetts Institute of Technology, Cambridge, MA 02139, USA}
\begin{abstract}
These lectures, presented at TASI 2016: Anticipating the Next Discoveries in Particle Physics, provide an introduction to some key methods and tools of indirect dark matter searches. Topics covered include estimation of dark matter signals, thermal freezeout and related scenarios, potential effects of dark matter annihilation on the early universe, modeling photon signals from annihilation or decay, and a brief and qualitative introduction to diffusive propagation of cosmic rays. The second half of the notes gives a status report (circa summer 2016) on selected experimental searches, the resulting constraints and some potential signal candidates. These notes are intended as an introduction to indirect dark matter searches for graduate students, focusing on back-of-the-envelope estimates and useful concepts rather than detailed quantitative computations.
\end{abstract}

\preprint{MIT-CTP/4946}

\maketitle

\tableofcontents

\newpage

\section{Motivations and estimates for indirect searches}

The dark matter (DM) searches falling under the umbrella of \emph{indirect detection} seek to identify possible visible products of DM interactions, originating from the DM already present in the cosmos. In particular, indirect searches often focus on searching for Standard Model (SM) particles produced by the decay or annihilation of dark matter, or their secondary effects. Indirect detection benefits from the huge amount of ambient DM (with energy density five times that of baryonic matter, over cosmological volumes), and the existence of  telescopes -- originally designed to answer other questions in astronomy and astrophysics -- that provide sensitivity to exotic sources of SM particles, especially photons, over an enormous range of energies. However, indirect searches face challenges because DM is known to interact only weakly with the SM, so the rate of particle production is expected to be small, and many possible detection channels have large potential backgrounds from astrophysical particle production.

Nonetheless, indirect detection has the potential to probe questions whose answers are much more indirectly linked to observable quantities, or even wholly inaccessible, in direct or collider-based searches. Two such key questions are:
\begin{enumerate}
\item Is DM perfectly stable?
\item What is the explanation for the observed abundance of DM?
\end{enumerate}

\subsection{Decaying dark matter}

In many theoretical models, the DM candidate is rendered stable by some unbroken symmetry, with the classic example being R-parity in supersymmetric theories. In this setup, superpartners are all odd under R-parity, whereas SM particles are all R-even. Consequently the lightest superpartner (LSP) cannot decay, since final states involving other superpartners are kinematically forbidden, and final states involving only SM particles would violate R-parity. This is an advantageous feature of such models, since in order for DM to be present in large quantities today (without having an enormously larger abundance at the epoch of recombination, violating limits from the cosmic microwave background), the lifetime $\tau$ of the DM must be longer than the age of the universe: $\tau \gg 10^{10}$ yr $\sim$ few $\times 10^{17}$ s. 

However, symmetries of a low-energy theory are often broken by some high-scale physics; if this is the case, then since the symmetry-breaking operators are suppressed at low energies, the timescale for decay through these operators can be extremely long. 

Suppose first, for example, that a $\mathcal{O}$(TeV) DM candidate decays through a higher-dimensional operator suppressed by the GUT scale, $\sim 2\times 10^{16}$ GeV. If the operator is dimension-5, this corresponds to a lifetime:
\begin{equation}\tau \sim M_\text{GUT}^2/m_\text{DM}^3 \sim (2\times 10^{16} \text{GeV})^2/(10^3 \text{GeV})^3 \sim 4 \times 10^{23} \text{GeV}^{-1} \sim 1 s. \end{equation}
This lifetime is clearly far too short. Even a suppression at the Planck scale would only increase the lifetime associated with such a dim-5 operator by $\sim 6$ orders of magnitude, leading to a lifetime less than a year (1 year $\sim 3\times 10^7$ s).

But suppose we consider instead a dimension-6 GUT-suppressed operator. Then for the same parameters as above,
\begin{equation}\tau \sim M_\text{GUT}^4/m_\text{DM}^5 \sim (2\times 10^{16} \text{GeV})^4/(10^3 \text{GeV})^5 \sim 10^{50} \text{GeV}^{-1} \sim 10^{26} s. \end{equation}

This lifetimes is $\sim 9$ orders of magnitude longer than the age of the universe, so there would be no observable change in the overall DM abundance due to these decays. With a decay lifetime longer than the age of the universe, DM would never decay to produce visible particles within the short timescale probed at a collider experiment; any DM produced would escape the detector before decaying. But if DM is a new particle, then the number of DM particles in our Galaxy is enormous, and some of them will be decaying at any given time: as we will see, if such a decay with a lifetime around $10^{26} s$ produced SM particles, those particles could themselves be observable.

Let us now estimate how many visible particles would be produced by such DM decays, and could potentially be observed. Consider a volume element $dV$ at a distance $r$ from Earth, and suppose the DM number density in that volume element is $n$; then assuming the decay lifetime $\tau$ is much longer than the age of the universe, the rate of decays from that volume element per unit time is $n dV/\tau$. If the system is in a steady state, and each decay produces one observable particle, then a detector of area $A$ a distance $r$ from $dV$ will measure $dN/dt = (A/4\pi r^2) \times n dV/\tau$ particles. Writing $dV = r^2 \sin\theta d\theta d\phi dr$, we have $dN/dt = A n(\vec{r}) (d\Omega/4\pi) dr/\tau$, and the total observed signal will be obtained by integrating over the source volume.

For a quick estimate, considering the signal from our local halo, let's take $n(\vec{r})$ to be the number density of DM at Earth, within a 1kpc radius of Earth -- this is roughly the propagation distance of weak-scale electrons and positrons. (At greater distances, those electrons and positrons will have lost the bulk of their energy, which can make them difficult to observe.) Thus the total number of counts/unit time becomes $dN/dt = A n r /\tau = A (0.4 \text{GeV/cm}^3) 1 \text{kpc}/(m_\text{DM} \tau)$, where we have used the fact that in the neighborhood of the Earth, the local DM mass density is $\rho \sim 0.4$ GeV/cm$^3$ (e.g. Ref.~\cite{Read:2014qva} and references therein). As a note, the cosmological critical density is $\rho \sim 5 \times 10^{-6}$ GeV/cm$^3$, and consequently the cosmological DM density is $\rho \sim 10^{-6}$ GeV/cm$^3$. 
 
Using $m_\text{DM} = 1$ TeV and $\tau = 10^{26} s$, as in our benchmark case above, we obtain $dN/dt = 10^{-4}$/s for a 1m$^2$ detector, or a few thousand events/year. So this suggests we should be able to see signals of electrons and positrons produced by TeV-scale DM decaying through a dimension-6 GUT-suppressed operator, with detectors of order a square meter in size. A benchmark detector of this kind is AMS-02.\footnote{\texttt{http://www.ams02.org/}} AMS-02, and the earlier detectors PAMELA and the Fermi Gamma-Ray Space Telescope (hereafter \emph{Fermi}), have observed a rise in the fraction of cosmic-ray positrons at energies above $\sim 10$ GeV \cite{Adriani:2008zr, FermiLAT:2011ab, Aguilar:2013qda}; DM decay is one possible explanation\citep{Arvanitaki:2009yb}, although it is increasingly in tension with constraints on DM decay from observations of photons from various regions of the sky (e.g. Refs.~\cite{Dugger:2010ys, Cohen:2016uyg}).

\begin{table}
	\begin{center}
	\begin{tabular}{cc}
	\begin{tikzpicture}
		  \begin{feynman}
			\vertex (a1) {\(\nu_S\)};
			\vertex[right=1cm of a1] (a2);
			\vertex[above right=1cm and 2cm of a2] (c1);
			\vertex[below right=1cm and 2cm of a2] (b1);
			\vertex[right=1cm of c1] (c2)  {\(\gamma\)};
			\vertex[right=1cm of b1] (b2) {\(\nu\)};
			\diagram*{
				(a2) -- [fermion,  edge label'=\(l^-\)] (c1);
				(c1) -- [fermion,  edge label'=\(l^-\)] (b1);
				(a2) -- [boson, edge label'=\(W^+\)] (b1);	
				(a1) -- [fermion] (a2);
				(b1) -- [fermion] (b2);
				(c1) -- [boson] (c2);		
			};
		  \end{feynman}
		\end{tikzpicture} & 
		\begin{tikzpicture}
		  \begin{feynman}
			\vertex (a1) {\(\nu_S\)};
			\vertex[right=1cm of a1] (a2);
			\vertex[above right=1cm and 2cm of a2] (c1);
			\vertex[below right=1cm and 2cm of a2] (b1);
			\vertex[right=1cm of c1] (c2)  {\(\gamma\)};
			\vertex[right=1cm of b1] (b2) {\(\nu\)};
			\diagram*{
				(a2) -- [boson,  edge label'=\(W^+\)] (c1);
				(b1) -- [boson,  edge label'=\(W^+\)] (c1);
				(a2) -- [fermion, edge label'=\(l^-\)] (b1);	
				(a1) -- [fermion] (a2);
				(b1) -- [fermion] (b2);
				(c1) -- [boson] (c2);		
			};
		  \end{feynman}
		\end{tikzpicture}\\
	\end{tabular}
	\end{center}
	
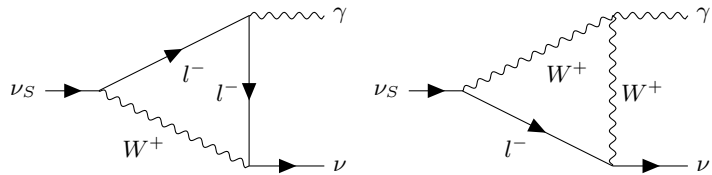
\captionof{figure}{Diagrams for radiative decay of a sterile neutrino, to a photon and a neutrino.}
	\label{fig:neutrino}
\end{table}

Another example of phenomenological interest is sterile neutrino decay; if a sterile neutrino exists and mixes with ordinary neutrinos, it can decay radiatively to a neutrino and a photon, as shown in Fig.~\ref{fig:neutrino}. Sterile neutrinos are another possible DM candidate. The lifetime of the sterile neutrino in this case is given by $\tau \sim 10^{30} s (10^{-7}/\sin^2(2\theta)) (1 \text{keV}/m_S)^5$, where $\theta$ is the mixing angle between the sterile neutrino and the active neutrino, and $m_S$ is the mass of the sterile neutrino (e.g. this relation is given in the review Ref.~\cite{Abazajian:2017tcc}). As we will discuss later in these lectures, the constraints on this decay rule out lifetimes shorter than $\sim10^{25-28} s$, depending on the DM mass.

It is interesting to note that since the rate of decays scales as number density/lifetime, the power injected by decays scales as $m_\text{DM} n /\tau = \rho/\tau$, and thus is essentially independent of the DM mass once the DM energy density has been measured. One result is that constraints on the DM decay lifetime are broadly similar over a very wide range of masses, as we will see in subsequent lectures. Since gravitational probes of DM are typically sensitive to the total DM mass inside a region, which also controls the total decay rate, constraints on the DM decay rate are also rather insensitive to uncertainties in the distribution of DM.

\subsection{Annihilating dark matter}

In many models, the late-time abundance of DM is controlled by its couplings to the SM. Exceptions can occur -- for example, in scenarios where the DM couples more strongly to other fields than the SM, the DM can undergo number-changing interactions that do not involve SM fields. Alternatively, a non-trivial initial condition that is not altered by subsequent interactions can fix the late-time DM abundance (e.g. if the DM possesses some matter-antimatter asymmetry, or simply has very weak interactions). But if DM self-annihilates and was once in thermal equilibrium with the SM, then in a large class of models, the late-time relic abundance can be entirely determined by the cross section for the annihilation process. This is typically called a ``thermal relic'' scenario; in such a scenario, measuring the annihilation cross section at late times by observing the annihilation products would give us a direct insight into the mechanism governing the DM abundance.

\begin{table}
	\begin{center}
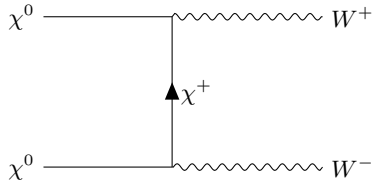

	\begin{tikzpicture}
		  \begin{feynman}
			\vertex (a1) {\(\chi^0\)}; 
			\vertex[below=2cm of a1] (a2) {\(\chi^0\)};
			\vertex[right=2cm of a1] (b1);
			\vertex[right=2cm of a2] (b2);
			\vertex[right=2cm of b1] (c1) {\(W^+\)};
			\vertex[right=2cm of b2] (c2) {\(W^-\)};
			\diagram*{
				(b2) -- [boson] (c2);
				(c1) -- [boson] (b1);	
				(a1) -- (b1);
				(a2) -- (b2);
				(b2) -- [fermion, edge label'=\(\chi^+\)] (b1);		
			};
		  \end{feynman}
		\end{tikzpicture}
	\end{center}
	\captionof{figure}{Tree-level annihilation of a supersymmetric wino $\chi^0$ to W bosons, through exchange of a chargino $\chi^+$.}
	\label{fig:wino}
\end{table}

Let us now consider some simple parametrics for such thermal relic models, and DM annihilation more generally. DM annihilation is typically not forbidden by whatever symmetry keeps the DM stable; e.g. for the LSP example discussed above, two DM particles both have R-parity -1, and so can annihilate to a pair of SM particles without violating R-parity. Annihilation thus often occurs at tree-level; it is not suppressed by any high scale. (For an example in a supersymmetric model, see Fig.~\ref{fig:wino}.) If there is no high-scale suppression, the annihilation cross section for DM coupling to its annihilation products with strength $\alpha_D$ can be generically estimated as $\sigma v_\text{rel} \sim \alpha_D^2/m_\text{DM}^2$, where $v_\text{rel}$ is the relative velocity between DM particles and $m_\text{DM}$ is their mass, although this expression (as we will discuss) is not valid for all models.

The rate of annihilations per unit volume per unit time, for two distinguishable annihilating particles with number densities $n_1$ and $n_2$, is given by:
\begin{align} \sigma v_\text{rel} n_1 n_2 = & \sigma \text{(cross-section)} \times n_2 v_\text{rel} \text{(number density flux of particles} \nonumber \\
& \text{(incident on target)} \times n_1 \text{(number density of particles in target}) \end{align}
For identical annihilating particles, the same rate instead becomes $\sigma v_\text{rel} n^2/2$, where $n$ is the number density of the identical particles; the factor of 2 avoids double-counting, since there are only $N(N-1)/2$ distinct pairs given $N$ indistinguishable particles.

We see that the annihilation rate scales as $\rho^2/m_\text{DM}^2$ for fixed $\sigma$, and as $\rho^2/m_\text{DM}^4$ if $\sigma \propto 1/m_\text{DM}^2$ (which is natural if $m_\text{DM}$ is the only relevant mass scale). Likewise, the power injected by annihilation scales as $\rho^2/m_\text{DM}$ for fixed $\sigma$, or as $\rho^2/m_\text{DM}^3$ for $\sigma \propto 1/m_\text{DM}^2$. As a result, limits on annihilation signals generally become weaker as the DM mass increases.

\subsubsection{Estimating freezeout}

Let us now consider the simplest thermal relic scenario, where a single DM species annihilates to SM particles with velocity-independent $\sigma v_\text{rel}$. What does imposing the measured present-day relic density imply for annihilation cross sections?

In the case with no annihilation, the number of DM particles in a comoving volume would remain constant; if $n$ is the physical (not comoving) DM number density, we would have $\frac{d}{dt} (n a^3) = 0$, where $a$ is the scale factor. Expanding this out, we obtain $dn/dt + 3 (\dot{a}/a) n = 0$, i.e. $dn/dt + 3 H n = 0$ where $H = \dot{a}/a$ is the Hubble parameter.

In the presence of annihilation, the number density is additionally depleted, yielding:
\begin{equation} \frac{dn}{dt} + 3 H n =-  \frac{n^2}{2} \langle \sigma v_\text{rel} \rangle \times 2, \end{equation}
for indistinguishable annihilating particles; the second factor of 2 occurs because each annihilation removes two particles, and the use of $\langle \sigma v_\text{rel} \rangle$ rather than $\sigma v_\text{rel}$ indicates we are averaging over the DM velocity distribution. However, in the presence of annihilation to any set of non-DM particles, the inverse reaction -- non-DM particles colliding to produce DM particles -- can also occur in principle. Thus we can write:
\begin{equation} \frac{dn}{dt} + 3 H n =- n^2 \langle \sigma v_\text{rel} \rangle + \text{contribution from DM production}, \end{equation}
 If the DM particles and their annihilation products are in chemical equilibrium, then the contributions from the DM depletion and production processes should cancel out. Thus we can write the DM-production term as $n_\text{eq}^2 \langle \sigma v_\text{rel} \rangle$, where $n_\text{eq}$ is the number density of the DM when it is in chemical equilibrium with its annihilation products, so overall we have:
 \begin{equation} \frac{dn}{dt} + 3 H n = (n_\text{eq}^2 - n^2) \langle \sigma v_\text{rel} \rangle. \label{eq:boltzmann} \end{equation}
 If the DM is in equilibrium with the SM thermal bath, its equilibrium number density is given by $n_\text{eq} \sim (m_\text{DM} T)^{3/2} e^{-m_\text{DM}/T}$ when $T \ll m_\text{DM}$, and $n_\text{eq} \sim T^3$ where $T \gg m_\text{DM}$.
 
 When the annihilation rate goes to zero, $n$ evolves as $1/a^3$; when the annihilation rate is large, $n$ will be forced close to $n_\text{eq}$. The crossover between the two regimes occurs when $\langle \sigma v_\text{rel} \rangle n^2 \sim H n$, i.e. $n \langle \sigma v_\text{rel} \rangle \sim H$. Thus the DM density diverges from its equilibrium value, approaching the constant comoving density that we measure at late times, when the Hubble expansion time becomes comparable to the time needed for a given DM particle to annihilate; we refer to this point as ``freezeout''. 

A precise calculation of the eventual DM abundance requires numerically solving the evolution equation (eq.~\ref{eq:boltzmann}), accounting for the non-trivial temperature evolution of the SM thermal bath when the number of relativistic degrees of freedom is changing. An up-to-date treatment is given in Ref.~\cite{Steigman:2012nb}.

However, a simple estimate can be performed by neglecting these effects, and assuming that freezeout is abrupt and occurs when $H = n_\text{eq} \langle \sigma v_\text{rel} \rangle$, that $n$ tracks $n_\text{eq}$ up to this point, and that after this point the DM number density evolves proportionally to $a^{-3}$. We denote the temperature of the universe at freezeout by $T_f$; we also define the new variable $x \equiv m_\text{DM}/T$, and write $x_f \equiv m_\text{DM}/T_f$. Thus the late-time DM number density is given by $n_\text{today} \approx n_\text{eq}(T_f) a(T_f)^3/(a_\text{today})^3$.

There are broadly two cases to be considered; in the first case, the DM is a ``hot relic'', and freezes out while still highly relativistic. In this case $n_\text{eq}(T_f) a(T_f)^3$ is determined entirely by the number of degrees of freedom of the DM, and is almost independent of the freezeout temperature. Consequently, the late-time DM number density is also largely independent of the details of freezeout, and should be comparable to the number density of photons, since the number density of the DM was originally that of a relativistic species and it has only been diluted by the cosmic expansion, not by any other number-changing effects. (This conclusion may be evaded if there are large changes in the number of relativistic degrees of freedom coupled to DM and/or the SM between freezeout and the present day, that affect the two sectors differently, e.g. Ref.~\cite{Baek:2014poa}). This would suggest a DM mass around the eV scale, since the photon abundance is roughly $2\times 10^9 \times$ larger than the baryon abundance, the energy density in baryons is comparable to that in DM, and the mass of a proton is 1 GeV. In turn, this mass scale implies that the DM would be relativistic during the epoch relevant to structure formation, and would thus behave as ``hot DM''; a scenario where hot DM constitutes $100\%$ of the DM would lead to dramatic changes to structure formation, and is not consistent with observations (see e.g. Ref.~\cite{Primack:2000iq}, or Ref.~\cite{DiValentino:2015wba} for more recent constraints).

In the second case, freezeout occurs when the DM is non-relativistic. This scenario can naturally explain a large depletion in the DM number density relative to the photon number density, since at freezeout, $n_\text{eq} \sim (m_\text{DM} T_f)^{3/2} e^{-m_\text{DM} / T_F}$ is exponentially suppressed. 

Our condition for freezeout in this second scenario is thus that $H \sim (m_\text{DM} T_f)^{3/2} e^{-m_\text{DM} / T_F} \langle \sigma v_\text{rel} \rangle$. If we assume that freezeout occurs during the radiation-dominated epoch, $H^2 \propto \rho \propto T^4$, and thus $T \propto H^{1/2} \propto t^{-1/2}$ (ignoring any changes in the number of degrees of freedom contributing to the energy density of the universe). Within our approximations, we can thus write $H = H(T=m_\text{DM}) x^{-2}$, where as previously $x=m_\text{DM}/T$. Our freezeout criterion then becomes:
\begin{align} H(x=1) x_f^{-2} & \sim (m_\text{DM}^2)^{3/2} x_f^{-3/2} e^{-x_f} \langle \sigma v_\text{rel} \rangle, \nonumber \\
\Rightarrow e^{-x_f} & \sim \frac{x_f^{-1/2} H(x=1)}{m_\text{DM}^3 \langle \sigma v_\text{rel} \rangle}.\end{align}

Since the exponential scaling with $x_f$ on the LHS is much faster than the power-law scaling on the RHS, as a lowest-order approximation we can write $x_f \sim \ln(m_\text{DM}^3 \langle \sigma v_\text{rel} \rangle/H(x=1))$. Note that $x_f$ has only a logarithmic dependence on the DM mass and the annihilation cross section.

Ignoring changes in the number of degrees of freedom coupled to the thermal bath, the photon number density and DM number density will both scale as $1/a^3$ after freezeout, so for purposes of our estimate, we can approximate $\rho_\text{DM}/n_\gamma = m_\text{DM} n_\text{DM}/n_\gamma$ at freezeout, and require that this match the observed late-time value. Now by the same reasoning as above, $n_\text{DM}(x_f) \sim H(x=1) x_f^{-2}/\langle \sigma v_\text{rel} \rangle$, whereas $n_\gamma(x_f) \sim T_f^3 \sim m_\text{DM}^3/x_f^3$. Thus we can write:
\begin{equation} \frac{m_\text{DM} n_\text{DM}}{n_\gamma} (x_f) \sim \frac{H(T=m_\text{DM})}{m_\text{DM}^2} \frac{x_f}{\langle \sigma v_\text{rel} \rangle}.\end{equation}
Since $H\propto T^2$, $H(T=m_\text{DM})/m_\text{DM}^2$ is approximately independent of $m_\text{DM}$; writing $H \sim T^2/m_\text{Planck}$, we can write:
\begin{equation} \frac{\rho_\text{DM}}{n_\gamma} (x_f) \sim \frac{1}{m_\text{Planck}} \frac{x_f}{\langle \sigma v_\text{rel} \rangle}.\end{equation}
Since $x_f$ is roughly independent of $m_\text{DM}$ at lowest order, we see that fixing $\rho_\text{DM}/n_\gamma$ to its measured present-day value will also fix $\langle \sigma v_\text{rel} \rangle$, to a value that is approximately independent of the DM mass.

Now let us put in some numbers: the DM mass density is roughly $5\times$ the baryon mass density, or $\sim 5$ GeV $\times n_b \sim 5 \times n_\gamma \times 5 \times 10^{-10}$ GeV, since the baryon-to-photon ratio is $\sim 5\times 10^{-10}$. Since $m_\text{Planck} \sim 10^{19}$ GeV, we obtain:
\begin{align} & \rho_\text{DM}/n_\gamma \sim 3 \times 10^{-9} \text{GeV} \sim \frac{x_f}{\langle \sigma v_\text{rel} \rangle} 10^{-19} \text{GeV}^{-1} \nonumber \\
& \Rightarrow \langle \sigma v_\text{rel} \rangle \sim x_f \times 10^{-10.5} \text{GeV}^{-2}.\end{align}

Since $x_f$ is a log quantity, let us first guess that it is $\mathcal{O}(1)$, to give us an estimate of roughly what mass range will yield the correct cross section; if we take $\langle \sigma v_\text{rel} \rangle \sim \alpha_D^2/m_\text{DM}^2$, and choose $\alpha_D \sim 0.01$ to be comparable to the electroweak coupling of the SM, we infer that $m_\text{DM} \sim 10^3$ GeV should yield roughly the right relic abundance. Armed with this information, we can make a better estimate of $x_f$:
\begin{align} x_f & \sim \ln(m_\text{DM}^3 \langle \sigma v_\text{rel} \rangle/H(x=1)) \nonumber \\ & \sim \ln (m_\text{DM} m_\text{Planck} \langle \sigma v_\text{rel} \rangle) \nonumber \\ & \sim \ln (10^{22} \text{GeV}^2 \times 10^{-10.5} \text{GeV}^{-2}) \nonumber \\ & \sim 25. \end{align}
Substituting this back into our expression for $\langle \sigma v\rangle$ yields $\langle \sigma v_\text{rel} \rangle \sim 10^{-9} \text{GeV}^{-2} \sim 10^{-26}$ cm$^3$/s, and a natural mass scale for $m_\text{DM}$ around a few hundred GeV. 

A more careful calculation (e.g. Ref.~\cite{Steigman:2012nb}) gives $\langle \sigma v_\text{rel} \rangle \approx 2-3 \times 10^{-26}$ cm$^3$/s, almost independent of the DM mass; this cross section is known as the ``thermal relic'' cross section.

\subsubsection{Detectability of thermal relic DM}

Let us repeat the same calculation we performed for decay, but now for annihilation; where previously the rate of annihilation per unit volume per unit time was $n/\tau$, now it is $(n^2/2)\langle \sigma v_\text{rel} \rangle$. Making the same assumptions as previously (i.e. considering the signal from a sphere of uniform DM density and 1kpc radius surrounding the Earth), we see that the number of particles incident on a detector of area A per unit time is given by:
\begin{equation} \frac{dN}{dt} = \frac{A\langle \sigma v_\text{rel} \rangle}{2} (1 \text{kpc}) \frac{\rho^2}{m_\text{DM}^2} \sim 10^{-26} \text{cm}^3/\text{s} \times A \times (1 \text{kpc}) \times \left(\frac{0.4 \text{GeV}}{m_\text{DM}}\right)^2 \text{cm}^{-6}.\end{equation}

Choosing $A=1 \text{m}^2$ as previously, and $m_\text{DM} = 1$ TeV, this rate corresponds to $5 \times 10^{-8}/s$ or roughly one event per year. For a 100 GeV DM particle, the rate would be two orders of magnitude higher. As it turns out, 100 GeV DM annihilating with a thermal relic cross section is close to the sensitivity limit of several current indirect searches.

\subsubsection{Effects of DM annihilation in the early universe}

As well as searching directly for DM annihilation/decay products, we can explore their effects on the history of the universe. Let us examine the rate of DM annihilation per Hubble time, over the history of the cosmos, using thermal relic DM as a benchmark.

After freezeout, the DM number density scales as $n \propto a^{-3}$. The number of annihilations in a comoving volume $V_c$ in a Hubble time is thus given approximately by:
\begin{equation} N_\text{ann} \approx \frac{n^2 \langle \sigma v_\text{rel} \rangle}{2} V_c H^{-1} \propto a^{-6} \langle \sigma v_\text{rel} \rangle  a^3 H^{-1}.\end{equation}
Here we are assuming DM is its own antiparticle.

During radiation domination, $H^2 \propto \rho \propto a^{-4}$, so $H \propto a^{-2}$ and $N_\text{ann} \propto a^{-1}$ (assuming $\langle \sigma v_\text{rel} \rangle$ is constant, which is true for $s$-wave annihilation of non-relativistic particles to much lighter species). During matter domination, $H^2 \propto \rho \propto a^{-3}$, so $N_\text{ann} \propto a^{-1.5}$. During dark energy domination, $H$ is independent of $a$, and so $N_\text{ann} \propto a^{-3}$.

Note that the number of DM particles in a comoving volume stays constant over freezeout, so the scaling of $N_\text{ann}$ also describes the fraction of DM particles that annihilates in the course of a Hubble time. At freezeout, by definition a given DM particle annihilates on average once per Hubble time, i.e. a $\mathcal{O}(1)$ fraction of DM particles annihilates per Hubble time. 

Using these scaling relations, we can quickly estimate what fraction of the DM is annihilating at later times in the universe's history. For example, for 100 GeV thermal relic DM, freezeout occurs at a temperature of roughly 5 GeV. Big Bang Nucleosynthesis occurs at a temperature around 1 MeV, during the radiation-dominated epoch, so since the fraction of DM annihilating in a Hubble time scales as $1/a$, roughly one in 5000 DM particles will annihilate in a Hubble time during the epoch of BBN. The amount of power being injected by DM annihilation per Hubble time will thus be $\sim 10^{-3} \times$ the total mass density of baryons, or $\sim$ 1 MeV per baryon; this amount of energy injection has the potential to affect subdominant nuclear abundances during nucleosynthesis (see e.g. Ref.~\cite{Jedamzik:2009uy}, or Ref.~\cite{Poulin:2015opa} for a more recent study).

In the epoch of recombination, when the cosmic microwave background (CMB) radiation is released, the temperature of the photon bath is around 1 eV, and the universe has recently become matter-dominated. We thus expect the fraction of DM annihilating per Hubble time to drop by roughly a factor of $5 \times 10^9$ between freezeout and recombination, for 100 GeV DM (or slightly more, due to the last part of the evolution occurring in the matter-dominated regime). This corresponds to roughly $10^{-9}$ of the total baryon energy density, or 1 eV of energy per baryon, being liberated by DM annihilation per Hubble time. Since a single hydrogen ionization requires 13.6 eV of energy, this in turn implies that the annihilation of 100 GeV thermal relic DM has the power to ionize roughly 10\% of the hydrogen in the universe! 

Recombination is characterized by a sharp drop in the ambient ionization level and a corresponding increase in the amount of neutral hydrogen; an increase in the post-recombination ionization level by 10\% would be very visible, as the extra free electrons would provide a screen to the photons of the CMB. Measurements of the CMB are currently sensitive to changes of just a few $\times 10^{-4}$ in the ionization fraction during the cosmic dark ages after recombination (e.g. Ref.~\cite{Galli2009}).

A more careful calculation, taking into account the temperature changes of the photon bath due to the changing number of relativistic degrees of freedom, the evolution of the DM annihilation rate after matter-radiation inequality, the exact temperature of the universe during and after the cosmic dark ages, the presence of recombination as well as ionization, and the fraction of injected power proceeding into ionization, finds a smaller signal in the CMB, by a couple of orders of magnitude. This still places interesting parameter space within the reach of current CMB experiments; thermal relic DM with a velocity-independent $\sigma v$ can be ruled out below mass scales of a few tens of GeV, depending on the annihilation channel \cite{Slatyer2015a}.

To estimate the fraction of DM annihilating in the present day, we note that $a$ increases by roughly 9 orders of magnitude between the freezeout of a 100 GeV thermal relic and matter-radiation equality, and then another four orders of magnitude between matter-radiation equality and the present day, translating to $9+6=15$ orders of magnitude change in the fraction of DM annihilating (or a little more once the late-time dominance of dark energy is accounted for). So DM annihilation is rare today; outside bound structures, one would expect only one in $\sim 10^{15}$ DM particles annihilates in a Hubble time. Of course, in the present day we have galaxies and galaxy clusters, where the DM density can be much higher. As discussed above, the DM density in the neighborhood of the Earth is roughly $4 \times 10^5$ orders of magnitude higher than the cosmological density, and thus we would expect a few in $10^{10}$ particles to annihilate per Hubble time. (We can alternatively simply calculate $n \langle \sigma v_\text{rel} \rangle$ for a number density of (0.4 GeV/$m_\text{DM}$)/cm$^3$ and a cross section of $3 \times 10^{-26}$ cm$^3$/s, and compare to the lifetime of the universe, $\tau \sim 4 \times 10^{17}$ s; this gives us a rate of a few in $10^{11}$ particles annihilating per Hubble time, for 100 GeV DM.) 

Clearly, DM annihilation is not expected to deplete the DM content of our Galactic halo anytime soon! This calculation is also illustrative for understanding the self-interaction cross sections required for DM-DM scatterings to impact the small-structure of halos (e.g. Ref.~\cite{Spergel:1999mh}); typically, an average DM particle in the halo must interact at least once in the dynamical time of the halo for self-interactions to have a substantial effect. From the calculation above, we see that even if the dynamical time of the halo approaches the Hubble time, the required self-interaction cross section would need to be 10-11 orders of magnitude above the thermal relic cross section, for our benchmark of 100 GeV DM.

\subsection{Alternative paradigms}

There are many other possible answers to the question of how the correct DM abundance is generated, beyond the one we have presented above. An incomplete list of examples includes:
\begin{itemize} 
\item The DM may never have had appreciable interactions with the SM; its abundance may be entirely set by initial conditions (e.g. the misalignment mechanism for axion DM, see Ref.~\cite{Duffy:2009ig} for a review). I will not discuss axion DM in these lectures as it will be covered as a separate topic.
\item If the DM is asymmetric, i.e. DM and anti-DM are distinct particles with different abundances (anti-DM is taken to be less abundant, without loss of generality), then if the annihilation cross section is sufficiently large, the annihilation process will freeze out due to a lack of anti-DM before it would decouple in the symmetric case. This requires an annihilation cross section larger than the thermal relic value we derived above, but once this requirement is satisfied, the final DM abundance is set by the asymmetry rather than the annihilation cross section. This is analogous to how the relic abundance of ordinary matter is determined.

It is worth noting that the annihilation rate need only be slightly larger than the thermal relic value for this scenario to work, and for the anti-DM to be depleted to a level far below the relic density \cite{Graesser:2011wi}. In the standard thermal freezeout scenario every DM particle is annihilating against a target whose density is also Boltzmann-suppressed, but this is not true for anti-DM in an asymmetric scenario, since the DM abundance converges toward its minimum value set by the asymmetry, rather than toward zero. Thus the remaining anti-DM has many available targets to annihilate on, and is efficiently depleted by even modest annihilation rates.

In the simplest models, there is no indirect detection signal at late times as there is no anti-DM left and so annihilations cut off completely.  However, if the anti-DM population can be regenerated at some later time (e.g. Refs.~\cite{Falkowski:2011xh,Cirelli:2011ac,Tulin:2012re}), there can potentially be very large indirect-detection signals due to the large annihilation cross section. A comprehensive review of asymmetric DM is given in Ref.~\cite{Petraki:2013wwa}.
\item In the ``freeze-in'' class of scenarios, the initial abundance of the DM is small, and the DM is never completely in thermal equilibrium with the SM. DM-SM interactions act to produce DM rather than depleting it, until they freeze out; thus larger cross sections lead to higher DM abundance \cite{Hall:2009bx}. 
\item Either SM particles or states within an expanded ``dark sector'' may be produced at early times, and later decay to generate the DM; if the dark-sector states are only slightly heavier than the DM itself, they may be accessible by collisions at late times, leading to interesting phenomenology.
\end{itemize}

Even within the simple thermal relic scenario discussed above, changes to the velocity dependence of $\sigma$ can have marked effects on the late-time signatures while leaving freezeout largely unaffected. For example, if annihilation from an initial state with total orbital angular momentum $L=0$ ($s$-wave annihilation) is strongly suppressed, then the dominant annihilation during freezeout can occur from $L \ge 1$ initial states; the contributions to annihilation from such states scale as $\sigma v_\text{rel} \propto v_\text{rel}^{2L}$, at leading order in velocity. For $p$-wave annihilation ($L=1$), the most common example, $\sigma  v_\text{rel}$ is suppressed by $v_\text{rel}^2 \sim T/m_\text{DM}$, which is a factor of $\mathcal{O}(1/20)$ at freezeout as discussed above, but $\mathcal{O}(10^{-6})$ in the present-day Galactic halo, where the typical velocity of DM particles is $v\sim 10^{-3} c$. 

A good summary of when the $s$-wave contribution is absent or suppressed is given in Ref.~\cite{Kumar:2013iva}. For one example, consider Majorana fermion DM annihilating to SM scalars (i.e. Higgs fields). Because the particles in the initial state are identical fermions, their overall wavefunction must be antisymmetric. If their total orbital angular momentum $L=0$, the spatial part of the wavefunction is symmetric, so the spin configuration must be antisymmetric; i.e. they must be in the spin-singlet state with $S=0$. By angular momentum conservation, the final state must have $J=0$, and since the constituents are scalars with no spin, they must also have $L=0$. It follows that the initial state is CP-odd and the final state is CP-even, so the $s$-wave contribution must involve CP violation; if the CP violation is small, this contribution will be suppressed. 

Another classic example is the case of Majorana fermion DM annihilating through a $s$-channel $Z$ boson to light SM fermions $\bar{f} f$; again the initial state must have $S=0$ if it has $L=0$. If these interactions do not violate CP, then the final state must also have $L=0$ and hence $S=0$ by angular momentum conservation. Since the outgoing particles have opposite momenta and $S=0$ implies their spins point in opposite directions, their helicities must be equal. But the $Z$ boson only couples to left-handed fermions and right-handed antifermions, so in the limit where $m_f \rightarrow 0$, this amplitude must vanish. Consequently, the $s$-wave contribution to the cross section is suppressed by powers of $m_f/m_\text{DM}$.

An even more extreme version of low-velocity suppression occurs in the ``forbidden dark matter'' scenario \cite{DAgnolo:2015ujb}, where the relic density is controlled by annihilations that are kinematically suppressed in the late universe, e.g. because the DM annihilates to particles slightly \emph{heavier} than itself. Such processes are unsuppressed when the DM kinetic energy is much larger than the splitting between the DM mass and the mass of the annihilation products; this criterion can be easily satisfied at freezeout, when the kinetic energy is of order $m_\text{DM}/20$, and not be satisfied (except for the exponentially rare DM particles on the tail of the Boltzmann distribution) for kinetic energies of order $10^{-6} m_\text{DM}$.

In the other direction, long-range interactions between DM particles can enhance the annihilation cross section at low velocities, rather than suppressing it; this effect is called ``Sommerfeld enhancement'' (e.g. Ref.~\cite{Hisano:2004ds}). If the DM particles propagate in a long-range potential, approximated by a Coulomb potential with coupling $\alpha_D$, then the $s$-wave annihilation rate is enhanced by a factor $2\pi \alpha_D/v_\text{rel}$ for $v_\text{rel} \ll \alpha_D$. Higher partial waves, with angular momentum quantum number $L$ for the initial state, are enhanced by a factor of order $(\alpha_D/v_\text{rel})^{2L + 1}$ \cite{Cassel:2009wt}. Of course, the DM-DM interaction is likely not infinite range; the mass of the force carrier $m_A$ can be neglected when $m_A \lesssim m_\text{DM} v_\text{rel}$, but for $v_\text{rel} \lesssim m_A/m_\text{DM}$, the enhancement generally saturates. At specific values of $m_A/(\alpha_D m_\text{DM})$, resonances occur, and the $s$-wave enhancement can instead be enhanced by a factor proportional to $1/v^2$, down to the saturation velocity. If such Sommerfeld enhancements are present, the prospects for indirect detection of thermal relics can be greatly enhanced, especially in searches where the typical DM velocity is very low (e.g. limits from the cosmic dark ages, before the DM has formed into gravitationally bound structures). 

Temperature-dependent resonance effects can also increase the DM annihilation rate during freezeout without a commensurate increase in the late universe, or vice versa.

Within the general thermal freezeout framework, more wide-ranging changes are possible; the principles of the calculation are the same, but the Boltzmann equations are modified. For example, the DM may undergo number-changing interactions that also produce DM particles in the final state; these interactions may either involve only DM particles (e.g. $3\rightarrow 2$ reactions) or both DM and SM particles. The case of ``semi-annihilation'' \cite{DEramo:2010keq} corresponds to DM DM $\rightarrow$ DM + SM (or the SM particle may be replaced with another unstable particle), where ``DM'' here can denote different dark-sector states. In the case of e.g. $3\rightarrow 2$ annihilations \cite{Hochberg:2014dra}, the annihilation products are always mildly relativistic and so this process heats up the dark sector; couplings to the SM are required to cool down the DM, dissipating heat from the dark sector.

In coannihilation scenarios, at early times the DM may interact with, annihilate against or be partially comprised of another species, which no longer exists in the late universe (see e.g. Ref.~\cite{Baker:2015qna} for a recent discussion). The presence of this partner species during freezeout modifies the annihilation rate. For example, pure wino DM in supersymmetric models consists of a neutral Majorana fermion, but there is a slightly heavier chargino state present in the spectrum of the theory; during freezeout, both are present, as the wino-chargino mass splitting is small compared to the temperature at freezeout. Consequently the relic abundance of both the wino and chargino species needs to be computed (including annihilations that involve both winos and charginos simultaneously), and added together; at late times the chargino decays to the wino. The presence of such additional states in the early universe can either decrease or increase the late-time DM annihilation cross section relative to expectations from the simple thermal relic scenario, depending on the relative sizes of the various relevant annihilation rates.

In summary, there are many variations on the simple thermal relic scenario that lead to a large range of present-day annihilation signals. But the simple scenario ties the relic abundance directly to the annihilation cross section, and thus gives us a benchmark for which to aim. 

\section{Tools and guidelines for indirect searches}

Having developed some simple estimates for the annihilation/decay rates we can probe, and the rates we would expect from theoretical principles, let us now discuss in more detail what ingredients go into the signals that indirect DM searches seek to observe.

In principle DM annihilations or decays (or other processes) could produce any SM particles. Most of those SM particles will subsequently decay on short timescales. The signatures we can hope to detect are the stable particles at the end of those decay chains: electrons, positrons, protons, antiprotons, photons and neutrinos.

\subsection{Neutral particles}

\subsubsection{J-factors}

Let us begin with photons and neutrinos, which travel to us in straight lines. In general we have only a two-dimensional view of the sky, although in some circumstances we can discern the distance at which a particular photon was emitted, e.g. because we have redshift information. Thus what we observe, in general, will be the number of photons or neutrinos arriving at our detector from within a particular solid angle on the sky, within a particular time interval.

As previously, let us suppose our telescope/detector has area $A$, and consider the signal arising from a volume $dV$ located at coordinates $(r,\theta,\phi)$, where the Earth is at $r=0$. Suppose each annihilation/decay produces an energy spectrum of photons (or neutrinos) $\left(\frac{dN_\gamma}{dE}\right)_0$. If the energy of the photons/neutrinos does not change between production and reception (i.e. redshifting, absorption etc are negligible), then the spectrum of photons received at Earth per volume per time is given by:
\begin{equation} \frac{dN_\gamma}{dE dt dV} = \left(\frac{dN_\gamma}{dE}\right)_0 \frac{A}{4\pi r^2} \times \begin{cases} 
      \frac{1}{2} \langle \sigma v_\text{rel} \rangle n(\vec{r})^2 & \text{annihilation} \\
      \frac{n(\vec{r})}{\tau} & \text{decay}
   \end{cases}\end{equation}
Integrating along the line of sight, we find:
\begin{equation} \frac{dN_\gamma}{dE dt d\Omega} = \frac{A}{4\pi} \left(\frac{dN_\gamma}{dE}\right)_0 \times \begin{cases} 
      \frac{\langle \sigma v_\text{rel} \rangle}{2 m_\text{DM}^2}  \int^\infty_0 \rho(\vec{r})^2 dr & \text{annihilation} \\
      \frac{1}{m_\text{DM} \tau} \int^\infty_0 dr \rho(\vec{r}) & \text{decay}
   \end{cases}\end{equation}
  If the source of annihilation/decay is localized, it often makes sense to integrate over the solid angle subtended by the object, to obtain the full signal from that source:
  \begin{equation} \frac{dN_\gamma}{dE dt } = \frac{A}{4\pi} \left(\frac{dN_\gamma}{dE}\right)_0 \times \begin{cases} 
      \frac{\langle \sigma v_\text{rel} \rangle}{2 m_\text{DM}^2}  \int dr d\Omega \rho(\vec{r})^2 & \text{annihilation} \\
      \frac{1}{m_\text{DM} \tau} \int^\infty_0 dr d\Omega \rho(\vec{r}) & \text{decay}
   \end{cases}\end{equation}
   
   Typically we separate the piece of this expression dependent on the particle physics from that entirely determined by the distribution of the DM mass density $\rho(\vec{r})$, which can be predicted from N-body simulations and/or measured by gravitational probes. The latter is called the ``J-factor'' of the source, and for annihilation can be defined as (note that there is more than one convention for the normalization in common use):
   \begin{equation} J_\text{ann} \equiv \frac{1}{8\pi} \int dr d\Omega \rho(\vec{r})^2,\end{equation}
   so that $\frac{1}{A} \frac{dN_\gamma}{dE dt} = \frac{\langle \sigma v_\text{rel} \rangle}{m_\text{DM}^2} \left(\frac{dN_\gamma}{dE}\right)_0 J_\text{ann}$.

There is an additional simplification that can be applied if the source is spherically symmetric. If $d$ is the distance from the Earth to the center of the source, $r$ is (as previously) the distance from the Earth to the point of annihilation, and we choose the $z$-axis of our coordinate system to point in the direction of the center of the source, then spherical symmetry of the source implies that $\rho(\vec{r})$ is in fact $\rho(\sqrt{r^2 + d^2 - 2 d r \cos\theta})$. The integral over $d\phi$ can then be performed immediately. If the source is the center of the Galaxy and we are working in Galactic coordinates $l$ and $b$, then it is helpful to note that $\cos\theta = \cos l \cos b$.

The J-factors of different sources characterize the relative size of their expected annihilation signals. Especially for regions close to the centers of halos, the J-factor can depend sensitively on the presumed density profile; a common choice is to model halos as following the Navarro-Frenk-White (NFW) profile \cite{Navarro:1995iw}, $\rho \propto r^{-1}/(1 + r/r_s)^2$, where now $r$ denotes distance from the center of the halo and $r_s$ is a characteristic scale radius. Under this assumption, the dwarf satellite galaxies of the Milky Way have J-factors in the neighborhood $J_\text{ann} \approx 10^{17-20}$ GeV$^2$/cm$^5$ \cite{Chiappo:2016xfs}; the region within 1 degree of the Milky Way's center has $J_\text{ann} \approx 10^{22}$ GeV$^2$/cm$^5$.

Naively we would thus expect the Galactic Center to be a more promising target for annihilation searches than dwarf galaxies. However, astrophysical backgrounds in the Galactic Center and the surrounding region are also much higher than in dwarf galaxies, so it can be more difficult to distinguish any potential signal from the background. Dwarf galaxies contain few baryons, so are relatively clean targets for indirect searches. The typical velocity of DM particles in dwarfs is also much smaller than in the Galactic Center; this can reduce signals in dwarfs in models where the annihilation is suppressed at low velocities (e.g. where $p$-wave annihilation dominates), or enhance it in models where the converse is true (e.g. models with Sommerfeld enhancement). The expected signal at the Galactic Center also depends strongly on the assumed model for the Milky Way's DM density profile; however, if a possible signal were observed, it would be possible to infer information about the DM density profile from the morphology of the signal.

The J-factors listed above include only the contribution from the smooth NFW density profile; in reality, the presence of small-scale substructure could potentially greatly increase $J_\text{ann}$. DM halos are thought to form by accretion of many smaller halos that formed at earlier times, and annihilation can be further enhanced in these small dense structures (because $\langle \rho^2\rangle \ne \langle \rho\rangle^2$, and the former is the relevant quantity for annihilation). The effect grows with the size of the host halo (as larger halos can contain more substructure), and for galaxy clusters could potentially give rise to a $\mathcal{O}(10^3)$ enhancement to $J_\text{ann}$ \cite{2011PhRvD..84l3509P,Gao:2011rf} (although more recent studies suggest a smaller enhancement \cite{Anderhalden:2013wd}). However, the size of this ``boost factor'' is highly uncertain, as models predicting large enhancements tend to have most of the annihilation power arising from subhalos well below the mass scale which can resolved in simulations, i.e. $10^{5-6}$ solar masses.

For the case of decay, substructure is irrelevant, and the signal size is controlled by $\int \rho(\vec{r}) dr d\Omega$. If the source is distant, so the distance from Earth to every point in the source is approximately equal at $r\approx R$, then this integral becomes approximately $\frac{1}{R^2} \int \rho(\vec{r}) dV = M/R^2$, where $M$ is the total mass of the source. Thus the strongest signals come from targets that have large total DM mass and are also relatively close; some of the strongest constraints arise from study of galaxy clusters.

So far we have assumed that the spectrum of neutrinos/photons produced by DM annihilation, followed by prompt decays of the annihilation products, propagates to Earth essentially un-distorted. This is a good approximation for neutrinos and gamma-rays from our own Galaxy and nearby systems. However, for more distant targets, we must consider redshifting, and possibly absorption.

For a simple example, let us consider the isotropic signal from annihilation of DM in the intergalactic medium, with density equal to the overall cosmological DM density. Now we must integrate over photons (or neutrinos) originating from all possible redshifts; we are interested in the photon density and spectrum in a present-day volume $d V_0$, arising from annihilation at all earlier times. We can write:
\begin{equation} \frac{dN}{dE dV_0} = \int^0_\infty dz \frac{dt}{dz} \frac{dN_\gamma(z)}{dE} \frac{\langle \sigma v_\text{rel} \rangle}{2} n(z)^2 \frac{dV_z}{dV_0}\end{equation}
Here $d V_z$ is the physical volume at redshift $z$ corresponding to the same comoving volume as $dV_0$. Since $a=1/(1+z)$, $dV_z/dV_0 = (a/a_0)^3 = 1/(1+z)^3$. We are exploiting the fact that the photon number density is (in this case) the same everywhere in the universe, and the photon number per comoving volume is preserved under the cosmic expansion, to argue that the average photon number density originating from DM annihilation at redshift $z$ is depleted exactly by $(1+z)^3$ by the present day. Note also that $\frac{dN_\gamma(z)}{dE}$ on the right-hand side is the spectrum of photons produced by an annihilation at redshift $z$ which have energy $E$ \emph{today}. If we define $E_z = E (1+z)$, i.e. the energy of a photon at redshift $z$ if that photon has energy $E$ today, then $\frac{dN_\gamma(z)}{dE} = (1+z) \frac{dN_\gamma(z)}{dE_z} = (1+z) (dN_\gamma/dE^\prime)_0|_{E^\prime= E_z}$. The factor of $dt/dz$ is needed to convert the rate of annihilations per unit time into annihilations per change in redshift; note that $d/dt \ln(1+z) = -d/dt \ln a = - H(z)$, so $dt/dz = -1/(H(z) (1+z))$. Finally, the cosmological DM density $n(z)$ scales as $a^{-3}$, i.e. $(1+z)^3$.

Putting this all together, we obtain:
\begin{equation} \frac{dN}{dE dV_0} = \int^\infty_0 dz \frac{(1+z)^3}{H(z)} \rho(z=0)^2 \left[ \left. \left(\frac{dN_\gamma}{dE^\prime}\right)_0\right|_{E^\prime=E(1+z)} \frac{\langle \sigma v_\text{rel} \rangle}{2 m_\text{DM}^2} \right].  \end{equation}
The term in square brackets encapsulates the model-dependent particle physics. For decay, we would replace $\langle \sigma v_\text{rel} \rangle/2 m_\text{DM}^2$ with $1/m_\text{DM} \tau$, $\rho(z=0)^2$ with $\rho(z=0)$, and the $(1+z)^3$ factor with $(1+z)^0$; the result is otherwise the same.

As an example of using this result, consider annihilation to a pair of photons with energies equal to the DM mass, so $(dN/dE)_0 = 2 \delta(E - m_\text{DM})$. Then we have:
\begin{align} \frac{dN}{dE dV_0} & = \frac{\langle \sigma v_\text{rel} \rangle}{2 m_\text{DM}^2} \rho(z=0)^2 \int^\infty_0 dz \frac{(1+z)^3}{H(z)}  2 \delta(E (1+z) - m_\text{DM}) \nonumber \\
& = \frac{\langle \sigma v_\text{rel} \rangle}{2 m_\text{DM}^2} \rho(z=0)^2 \frac{(m_\text{DM}/E)^3}{H(z = m_\text{DM}/E - 1)}  \frac{2}{E}, \quad 0 \le E \le m_\text{DM} \nonumber \\
& = m_\text{DM} \langle \sigma v_\text{rel} \rangle \rho(z=0)^2 \frac{1}{H_0 \sqrt{\Omega_m (m_\text{DM}/E)^3 + \Omega_\Lambda}}  \frac{1}{E^4}, \quad 0 \le E \le m_\text{DM}, \end{align}
where in the last line we have neglected the contribution to $H$ from the radiation field, which is valid for small $z$.

In the general case, we will need to include both non-uniformity and redshifting, obtaining:
  \begin{equation} \frac{dN_\gamma}{dE dA dt } = \int \frac{d\Omega}{4\pi} \int dz \left. \left(\frac{dN_\gamma}{dE^\prime}\right)_0\right|_{E^\prime=E(1+z)}  \frac{1}{H(z) (1+z)^3} \times \begin{cases} 
      \frac{\langle \sigma v_\text{rel} \rangle}{2 m_\text{DM}^2} \rho(z,\theta,\phi)^2 & \text{annihilation} \\
      \frac{1}{m_\text{DM} \tau} \rho(z,\theta,\phi) & \text{decay}
   \end{cases}\end{equation}
   The special cases discussed above can be obtained by taking $\rho(z,\theta,\phi) = \rho(z=0)(1+z)^3$ on one hand (for the isotropic homogeneous case), or by setting $z=0$ and noting that $dz/H(z) = (1+z) d\ln(1+z)/H(z) = -(1+z) dt = -dt$ for $z=0$ (for the case where redshifting can be neglected), and then replacing $-dt$ with $dr$ for particles traveling toward Earth at lightspeed.
   
   To include absorption, we could include a factor of the form $e^{-\tau(E,z)}$ inside the integral, where the function $\tau(E,z)$ describes the optical depth for a photon emitted at redshift $z$ and with (measured at $z=0$) energy $E$. Non-uniform sources, redshifting and absorption can all be relevant when computing contributions to the ambient radiation fields from DM annihilation/decay over the history of the universe; see e.g. Ref.~\cite{Zavala:2011tt} for an example.

\subsubsection{The photon/neutrino spectrum}

Up to this point we have left the photon/neutrino spectrum per annihilation/decay completely unspecified. But our search strategy will depend on this spectrum, so let us now consider some possible options.

We often parameterize the possible photon/neutrino spectra in terms of the various possible 2-body SM final states, with the logic that if annihilation/decay to 2-body final states \emph{can} occur, then it will usually dominate the overall signal. This is not true in all cases -- for example, the annihilation to two particles may be $p$-wave and hence velocity-suppressed, whereas adding a third particle to the final state lifts the velocity suppression and allows $s$-wave annihilation \cite{Bell:2011if} -- but it is common in many models. Then we can consider the spectra of photon and neutrinos produced by DM annihilation to pairs of quarks, gauge bosons, leptons etc, and their subsequent decays, and test these spectra against observations.

Within these possible final states, there are three broad spectral categories for photons:
\begin{enumerate}
\item Hadronic / photon-rich continuum: if the DM annihilates to $\tau$ leptons, gauge bosons, or any combinations of quarks, then copious neutral and charged pions will be produced in the subsequent decays of those particles. Neutral pions decay to a photon pair ($\pi^0 \rightarrow \gamma \gamma$) with a 99\% branching ratio, so a broad spectrum of photons is produced, along with electrons and positrons from the charged pion decays.
\item Leptonic / photon-poor: the DM annihilation produces mostly electrons and muons.  Photons are produced directly only as part of 3-body final states, by final state radiation or internal bremsstrahlung; the rate for photon production is suppressed, and the photon spectrum is typically quite hard, peaked toward the DM mass \cite{Birkedal:2005ep}. (Note that similar hard photon spectra can be produced if the DM decays into a mediator that subsequently decays to photons, e.g. Ref.~\cite{Ibarra:2012dw}.) Copious charged leptons are produced.
\item Lines: the DM annihilates directly to $\gamma \gamma$ (or $\bar{\nu} \nu$ in the neutrino case), or a monoenergetic photon plus another particle. Such channels allow ``bump hunts'', and greatly reduce the possible astrophysical backgrounds; a clear detection of a gamma-ray spectral line would be very difficult to explain with conventional astrophysics. However, DM is known to carry no electric charge, and thus cannot couple directly to photons, so this signal must be suppressed at the 1-loop level at least, and is expected to be small.
\end{enumerate}

For neutrinos the qualitative picture is similar to photons for the hadronic and line cases (although searches for a neutrino line would be very experimentally challenging; on the other hand, the DM coupling to neutrinos need not be loop-suppressed), but the leptonic case can be different; annihilation to muons will produce an unsuppressed neutrino spectrum, whereas in the case of direct annihilation to electrons the neutrino signal would be expected to be very small.

It is possible, of course, that the DM does not annihilate directly into SM particles; if the DM annihilates to other particles in a dark sector, and these subsequently decay back to the SM (either directly or through some cascade), then the eventual photon spectra need not lie in the space spanned by the 2-body SM final states. Some discussion of the range of possible spectra and the implications for indirect detection can be found in Ref.~\cite{Elor2015a}.

Charged particles from DM annihilation can also give rise to \emph{secondary} photons, due to upscattering of ambient photons from starlight or the cosmic microwave background, and synchrotron radiation from high-energy charged particles propagating in a magnetic field. For leptonic channels, this is often the largest photon signal; however, it depends on modeling the propagation of the charged particles, which we will discuss shortly.

Astrophysical backgrounds for photon signals from DM vary depending on the photon energy, and hence on the DM mass. For sufficiently high-energy (gamma-ray) lines or sharply-peaked spectra, backgrounds are essentially non-existent; the only challenge is collecting sufficient statistics. For continuum gamma rays, cosmic rays interacting with the gas and starlight produce background photons -- hadron-hadron collisions produce neutral pions which decay to gamma rays, and cosmic rays upscatter ambient photons to gamma-ray energies. Pulsars also produce photons with energies of a few GeV and below. At X-ray energies, relevant for searches for sterile neutrino DM, there are continuum X-rays from hot gas, as well as spectral lines from various atomic processes. At radio and microwave energies, relevant for searches for synchrotron radiation from weak-scale DM annihilation products, backgrounds include the CMB, synchrotron radiation from conventional sources, and thermal emission from interstellar dust.

\subsection{Cosmic rays}
\label{sec:crs}

In contrast to photons and neutrinos, charged particles produced by DM annihilation diffuse through the Galactic magnetic fields rather than following straight-line paths; furthermore, they can lose energy rapidly, so even on sub-Galactic scales, their spectrum changes with distance from the source. Both the signal and the background are thus more theoretically challenging to model, and in the event of a possible signal being detected, there will be little spatial information as to the origin of the cosmic rays -- their directionality is washed out by the ambient magnetic fields.

Public tools for modeling cosmic-ray propagation numerically include DRAGON \cite{2010APh....34..274D, Evoli:2016xgn} and GALPROP \cite{galprop,Strong:1999sv}. Here I will briefly sketch the principles of cosmic-ray diffusive propagation underlying these codes, following the review of Ref.~\cite{Strong:2007nh}. Readers seeking a more detailed treatment may find it there.

 Let us write the number density of cosmic rays at a given energy as $dn_\text{CRs}/dE = \psi(\vec{x}, E, t)$. The evolution of this number density field is approximately governed by a diffusion equation:
 \begin{equation} \frac{\partial \psi}{\partial t} = D(E) \nabla^2 \psi + \frac{\partial}{\partial E} (b(E) \psi) + Q(\vec{x},E,t). \label{eq:diff} \end{equation} 
Here $D(E)$ is a diffusion coefficient, which we approximate to be independent of energy and time; $b(E)$ describes the energy losses for the cosmic-ray species in question; and $Q$ characterizes sources.

Other possible terms can be (and have been) added to this equation: convection of cosmic rays out of the Galactic plane can be described by a term of the form $\frac{\partial}{\partial z} (v_c \psi)$, where $v_c$ is the convection speed; decay or fragmentation of unstable cosmic rays can be described by a decay term $-\psi/\tau$; diffusive reacceleration corresponds to a term of the form $\frac{\partial}{\partial p} \left[ p^2 D_{pp} \frac{\partial}{\partial p} \left(\psi/p^2\right)\right]$. But for the purposes of this lecture we will only consider the simple form of eq.~\ref{eq:diff}.

In order to solve this equation, we need to impose boundary conditions. The standard approximation, which I will follow, is to treat the Galaxy as a cylindrical slab with height $h$ (of order a few kpc) and radius $R$ (of order a few $\times$ 10 kpc), and impose a free-escape condition at the slab boundaries. The diffusion coefficient is generally parameterized as $D(E) = D_0 (E/E_0)^\delta$, where $E_0 = 1$ GeV, $D_0 \sim$ few $\times 10^{28}$ cm$^2$/s. A number of different values for $\delta$ are in use, with $\delta \sim 0.3$ and $\delta \sim 0.7$ being common bracketing values. $\delta=1/3$ and $\delta=1/2$ correspond to theoretical scenarios called Kolmogorov-type and Kraichnan-type diffusion, respectively, corresponding to different spectra for the magnetic field turbulence; higher values of $\delta \sim 0.7$ were preferred by earlier cosmic-ray data, e.g. Ref.~\cite{2012A&A...539A..88C}, although the latest data seem to suggest a smaller value $\delta \sim 0.4-0.5$ \cite{Yuan:2017ozr}. The boundary parameters $R$ and $h$ and the diffusion parameters can be tuned to match measured cosmic-ray data.

The timescale for diffusion is given by $\tau_\text{diff} \sim R^2/D(E)$; the timescale for energy losses by $\tau_\text{loss} \sim E/b(E)$. If we assume a steady-state regime, $\partial \psi/\partial t = 0$, and we approximate $\partial/\partial E$ with $1/E$ -- which is approximately correct for power-law spectra -- and $\nabla^2 \psi$ as $\psi/R^2$, then we can rewrite the diffusion equation (eq.~\ref{eq:diff}) in the illustrative form:
\begin{equation} -\frac{\psi}{\tau_\text{diff}} - \frac{\psi}{\tau_\text{loss}} + Q \approx 0.\end{equation}
This equation has the approximate solution $\psi \sim Q \text{min}(\tau_\text{diff},\tau_\text{loss})$.

There are two limiting cases, the diffusion-dominated regime where $\tau_\text{diff} \ll \tau_\text{loss}$, and the cooling-dominated regime where $\tau_\text{loss} \ll \tau_\text{diff}$. In the diffusion-dominated regime where energy losses are slow, which is the relevant case for protons and antiprotons, we have $\psi \propto Q(E) E^{-\delta}$, where $Q(E)$ describes the source spectrum of the cosmic rays. Thus diffusion softens the injected spectrum by an index set by $\delta$. The observed spectrum of protons has $dn/dE \propto E^{-2.7}$; if the injected spectrum for Galactic cosmic rays is $dn/dE \propto E^{-2}$, characteristic of particles accelerated by strong shocks \cite{Baring:1997ka}, then this would suggest $\delta \sim 0.7$.

In the cooling-dominated or loss-dominated regime, energy losses are fast relative to diffusion; this is the relevant regime for high-energy electrons and positrons. The main energy loss processes are synchrotron radiation in ambient magnetic fields, and inverse Compton scattering of ambient photons. If the cosmic rays are not too energetic -- i.e. the geometric mean of their energy and the ambient photon energy is less than the electron mass -- then the energy loss rate for inverse Compton scattering has a simple form, $dE/dt \propto E^2$; this is also the case for synchrotron radiation. Thus in this case we can write $b(E) \propto E^2$, and $\tau_\text{loss} \propto E^{-1}$, vs $\tau_\text{diff} \propto E^{-\delta}$. For $0 < \delta < 1$, as a consequence, losses increasingly dominate ($\tau_\text{loss}$ is smaller) at higher energies. Thus we expect a spectrum of $Q(E) E^{-\delta}$ for low-energy electrons and positrons, breaking to $Q(E) E^{-1}$ at high energies.

As cosmic rays propagate through the Galaxy, protons can scatter on the ambient gas, producing secondary photons, positrons and electrons. For these secondary cosmic rays, $Q(E)$ should be replaced by the steady-state proton spectrum. If the original $Q(E) \propto E^{-2}$ for all species, and so the steady-state proton spectrum is $\propto E^{-(2 +\delta)}$, then the secondary positron spectrum should have a spectrum of $E^{-(2 + 2\delta)}$ at low energies, breaking to $E^{-(3 + \delta)}$ at high energies, in contrast to the primary positron spectrum, which is proportional to $E^{-(2+\delta)}$ at low energies and breaks to $E^{-3}$ at high energies. More generally, secondaries should have a softer spectrum than primaries by a factor of $E^{-\delta}$, if the primaries are in the diffusion-dominated regime.

Suppose that instead the source term is DM annihilation to $e^+ e^-$, with the electron and positron each having energy equal to the DM mass: $Q(E) = Q_0 \delta(E - m_\text{DM})$. In this case the approximation of the injected spectrum as a power law is clearly inaccurate. Let us consider the steady-state spectrum in the loss-dominated regime, so the diffusion equation becomes:
\begin{equation} \frac{\partial}{\partial E} (b(E) \psi) = -Q_0 \delta(E - m_\text{DM}).\end{equation}
Integrating both sides gives:
\begin{equation} \psi(E) = \frac{Q_0}{b(E)}, \quad 0 \le E \le m_\text{DM}.\end{equation}
So for $b(E) \propto E^2$ as discussed above, $\psi(E) \propto Q_0 E^{-2}, 0 \le E \le m_\text{DM}$; that is, the steady-state spectrum is a smooth power law with a sharp cutoff at the DM mass. Due to the relatively hard spectrum -- harder than one would expect from shock-accelerated cosmic rays softened by diffusion and/or losses -- combined with the sharp endpoint, it may in principle be possible to distinguish such a signal from background. But this is quite non-trivial, as astrophysical sources can also have energy cutoffs, and experimental energy resolution is a limiting factor.

\section{Current indirect searches for dark matter}

In the previous two lectures I have argued that DM annihilation (decay) at interesting cross sections (lifetimes) can have observable traces in the present day and over the history of the universe. The possible signals span a huge range of energies; we have discussed how to calculate the expected signals in photons, neutrinos and charged particles. I will now outline the status of current  searches for such signals.

\begin{figure}
\centerline{\includegraphics[width=10cm]{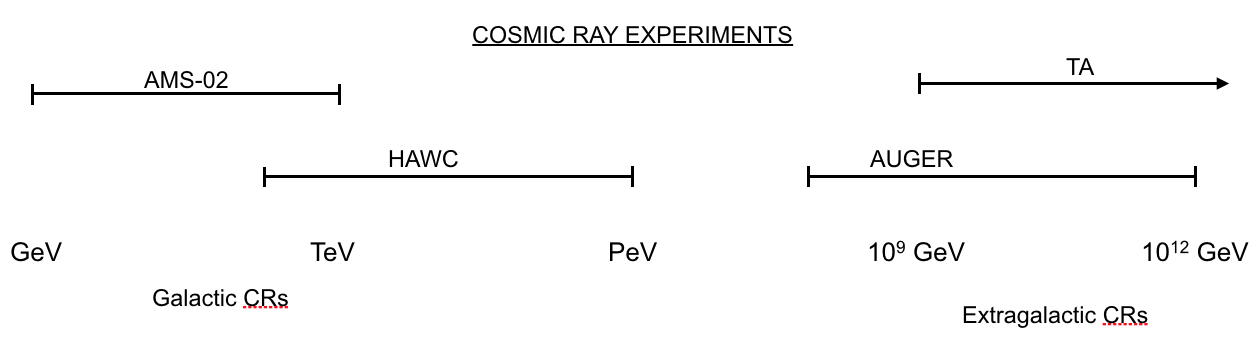}}
\vskip 1mm
\centerline{\includegraphics[width=6cm]{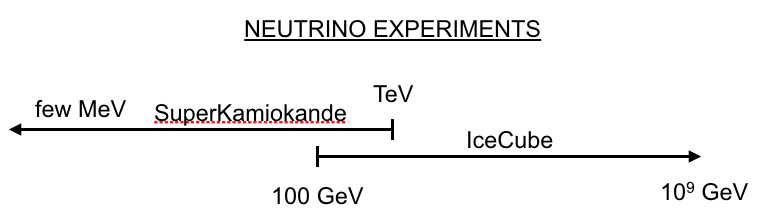}}
\vskip 5mm
\centerline{\includegraphics[width=8cm]{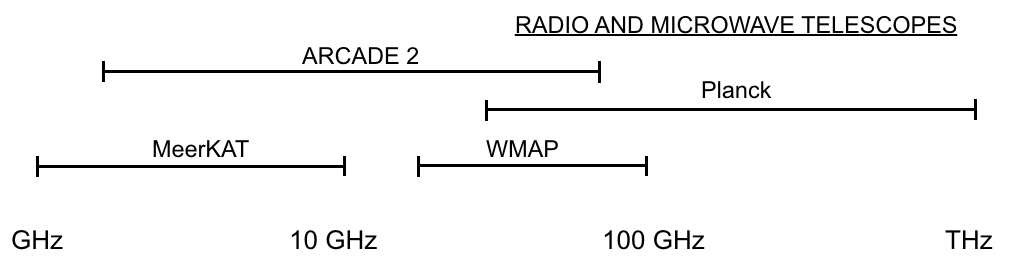}}
\vskip 5mm
\centerline{\includegraphics[width=11cm]{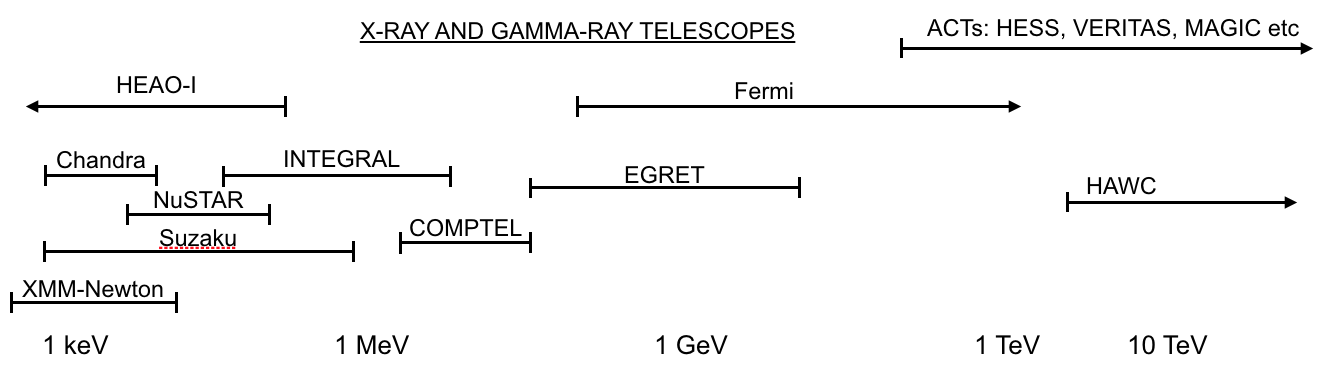}}
\caption{A sketch of the approximate energy ranges covered by a (incomplete) selection of present-day cosmic-ray, neutrino and photon telescopes. From top to bottom, the panels describe cosmic-ray detectors, neutrino telescopes, photon telescopes in the radio and microwave bands, and photon telescopes in the hard UV, X-ray and gamma-ray bands.}
\label{fig:currentexp}
\end{figure}

Fig.~\ref{fig:currentexp} summarizes the energy reach of several current telescopes that will be discussed below, spanning photon energies from radio to gamma-rays, as well as neutrino and cosmic-ray detectors.

\subsection{The cosmic microwave background}

As we discussed and estimated earlier, DM annihilation or decay during the cosmic dark ages can cause additional ionization of the ambient hydrogen gas \cite{Adams:1998nr, Chen:2003gz, Padmanabhan2005}; the resulting free electrons scatter the CMB photons and modify the measured anisotropies of the CMB. In order to calculate this effect in detail, we need the following ingredients:
\begin{enumerate}
\item The spectrum of stable electromagnetically interacting particles produced by the DM annihilation/decay, and the redshift dependence of the energy injection.
\item A calculation of how these electromagnetically interacting particles cool and lose their energy, what fraction of their energy is converted into hydrogen ionization, and how long the cooling process takes.
\item A calculation of how extra ionizing energy modifies the ionization history of the universe, and how modifications to the ionization history affect the anisotropies of the CMB.
\end{enumerate}

The third ingredient is available in public codes: RECFAST \cite{Seager:1999bc}, HyREC \cite{AliHaimoud:2010dx} and CosmoRec \cite{Chluba:2010ca} calculate the modified ionization history, while CAMB \cite{Lewis:2002nc} and CLASS \cite{2011arXiv1104.2932L} can translate arbitrary ionization histories into modifications to the CMB anisotropy spectra. The first ingredient can usually be calculated fairly straightforwardly once the DM model is determined; it is the same spectrum-at-source relevant to other indirect searches. The second ingredient has been calculated and tabulated in Ref.~\cite{Slatyer2015} for electrons, positrons and photons, for keV-multi-TeV injection energies; Ref.~\cite{Weniger2013} has performed a more limited calculation of the effect of protons and antiprotons and argues that their contribution to the ionization history will generally be small. 

Note that the second ingredient here is agnostic as to the origin of the electromagnetically interacting particles, and the third ingredient does not require knowledge of the source of the extra ionization. Thus details of the particle physics model enter only in the first ingredient; separating the ingredients in this way thus allows the calculations in (2) and (3) to be worked out for arbitrary injections of electromagnetically interacting particles, and then applied to specific DM models as needed.

It turns out that the limit on $s$-wave annihilating DM from the CMB depends on essentially one number: the excess ionization at redshift $z\sim 600$ \cite{Galli2009,Finkbeiner2012, Slatyer2015a}. For decay, the signal is similarly dominated by redshift $z\sim 300$ \cite{Slatyer:2016qyl,Poulin:2016anj}. The shape of the CMB anisotropies is nearly model-independent; this parameter fixes the overall normalization. We can thus define an efficiency factor $f_\text{eff}$ such that the signal in the CMB is directly proportional to $f_\text{eff} \langle \sigma v_\text{rel} \rangle/m_\text{DM}$ for ($s$-wave-dominated) annihilation, or to $f_\text{eff}/\tau$ for decay, where $f_\text{eff}$ is a model-dependent efficiency factor. Recall that $\langle  \sigma v_\text{rel} \rangle/m_\text{DM}$ ($1/\tau$) controls the rate of energy injection from annihilation (decay), as discussed earlier.

\begin{figure*}
\includegraphics[width=0.45\textwidth]{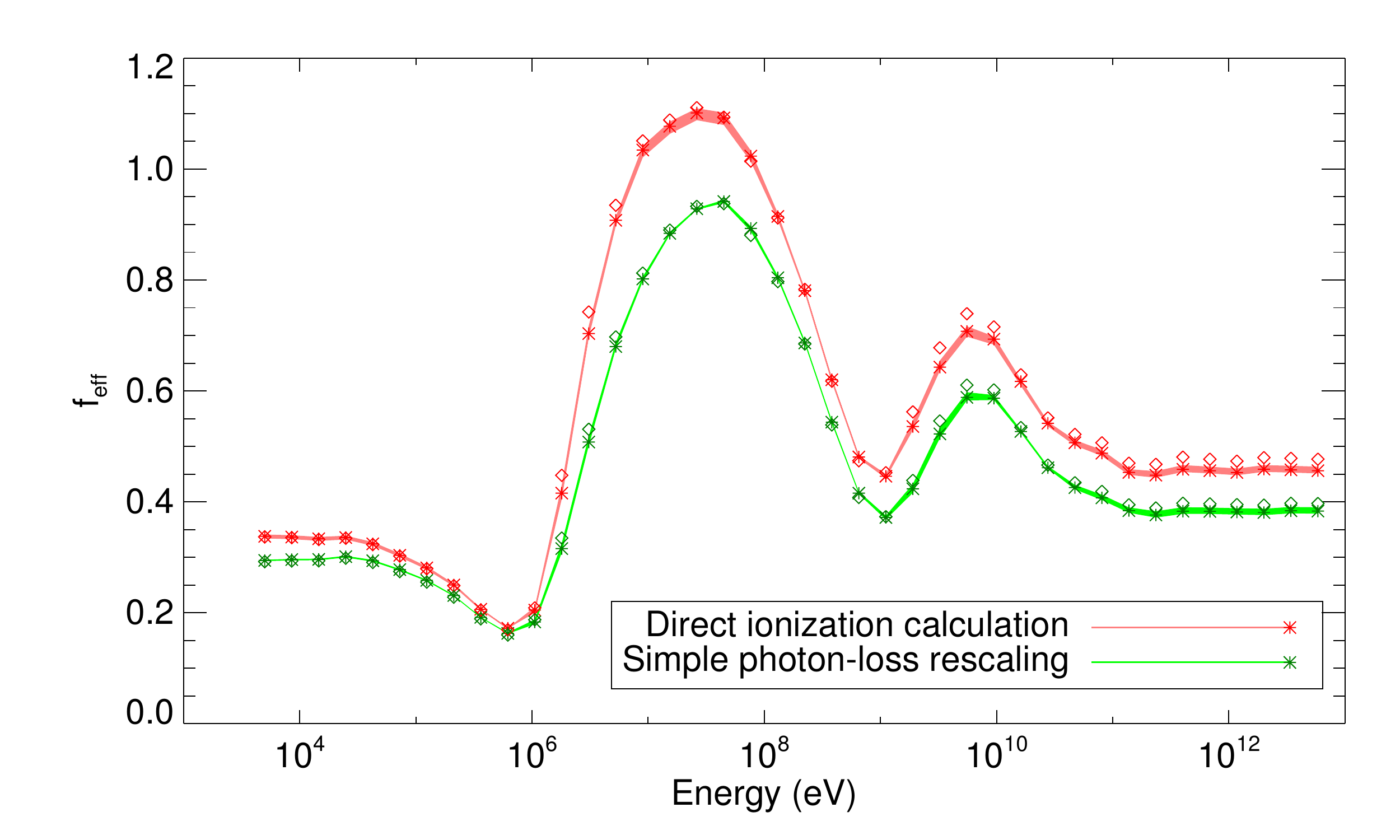}
\includegraphics[width=0.45\textwidth]{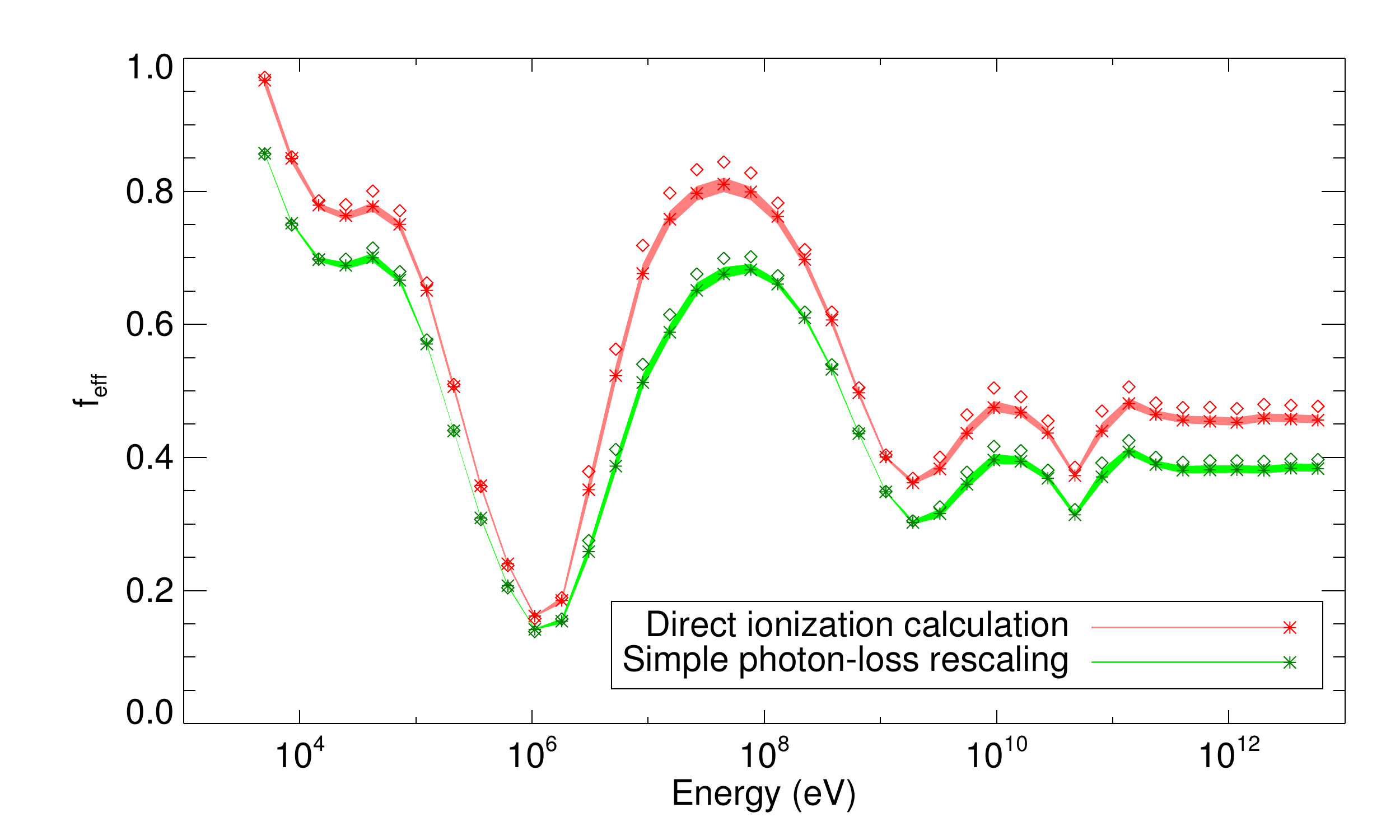}
\caption{\label{fig:feff}
$f_\mathrm{eff}$ coefficients as a function of energy for $e^+ e^-$ (left column) and photons (right column), for annihilating DM. The widths of the red and green bands indicate systematic uncertainties in the derivation of the $f_\text{eff}$ factors. Red and green stars indicate two different methods of calculating the efficiency factors, with the red points corresponding to the detailed calculation and the green to a simplified version. Diamonds indicate $f_\mathrm{eff}$ evaluated as the value of the efficiency of deposition at $z=600$, rather than taking into account the full redshift dependence. Figure reproduced from Ref.~\cite{Slatyer2015a}.
}
\end{figure*}

 \begin{figure*}\centerline{
   \includegraphics[width=6.5cm]{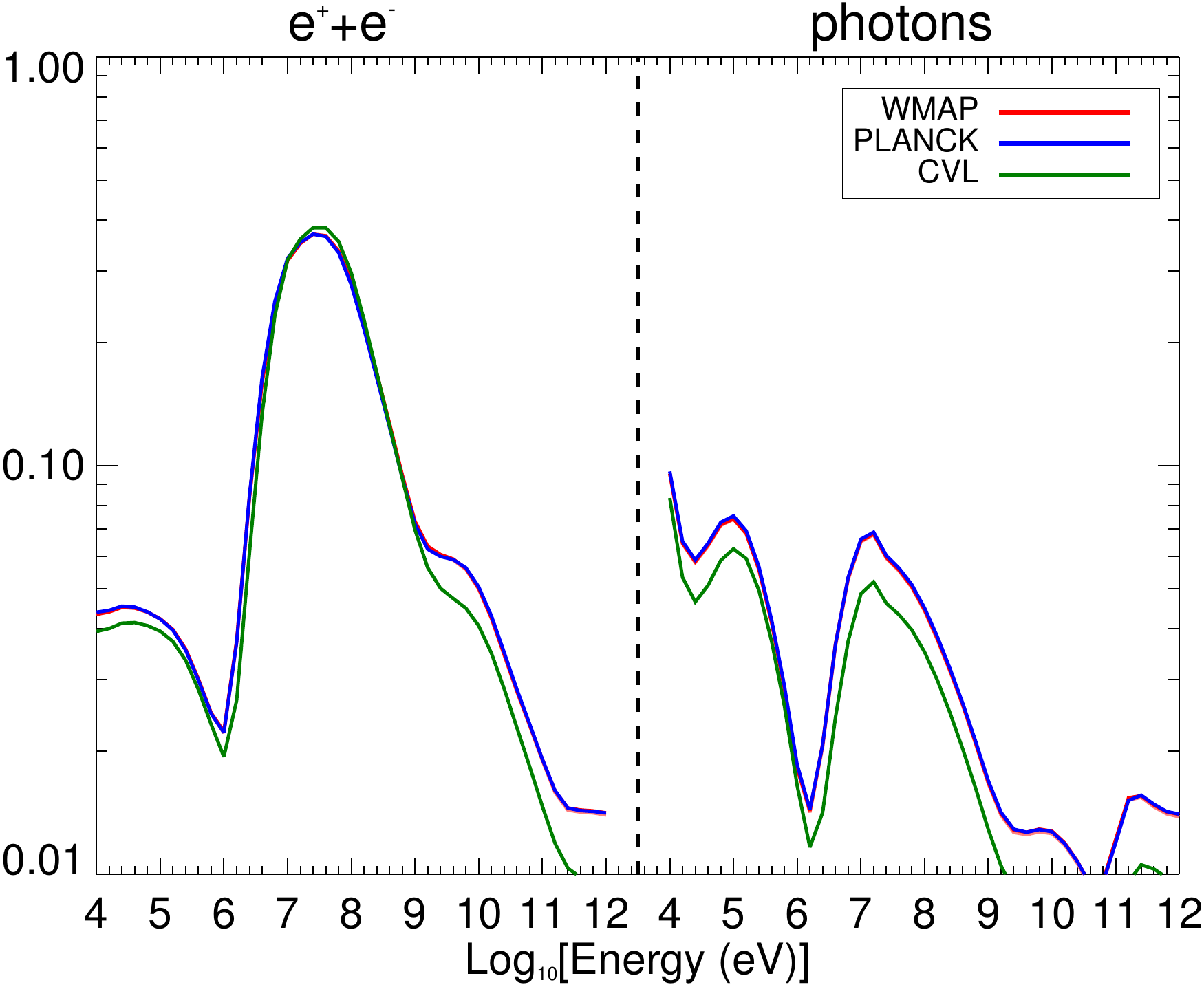}}
  \caption{$f_\mathrm{eff}$ coefficients as a function of energy for $e^+ e^-$ (left panel) and photons (right panel), for decaying DM with a lifetime much longer than the age of the universe. Figure reproduced from Ref.~\cite{Slatyer:2016qyl}.}
\label{fig:feffdec}
\end{figure*}

The parameter $f_\text{eff}$ depends primarily on how much of the injected power proceeds into electromagnetically interacting particles, as opposed to neutrinos. Secondarily, it depends on the spectrum of the injected electrons, positrons and photons; most of the variation occurs for particle energies below the GeV scale. Fig.~\ref{fig:feff} displays the numerically-computed $f_\text{eff}$ factors for photons and $e^+ e^-$ pairs injected at different energies, for the case of $s$-wave annihilation \cite{Slatyer2015a}; Fig.~\ref{fig:feffdec} shows the equivalent factors for decay with a lifetime longer than the age of the universe \cite{Slatyer:2016qyl}. Results for arbitrary photon/electron/positron spectra can be obtained by integrating the product of the spectrum with $f_\text{eff}(E)$, to obtain an average $f_\text{eff}$ value. Note that the normalization in these figures is arbitrary; having set a constraint on any one reference DM model, one can convert the bound to a limit on any other DM model, by using the relative $f_\text{eff}$ values for the reference model and the model of interest.

\begin{figure*}
\centerline{\includegraphics[width=0.8\textwidth]{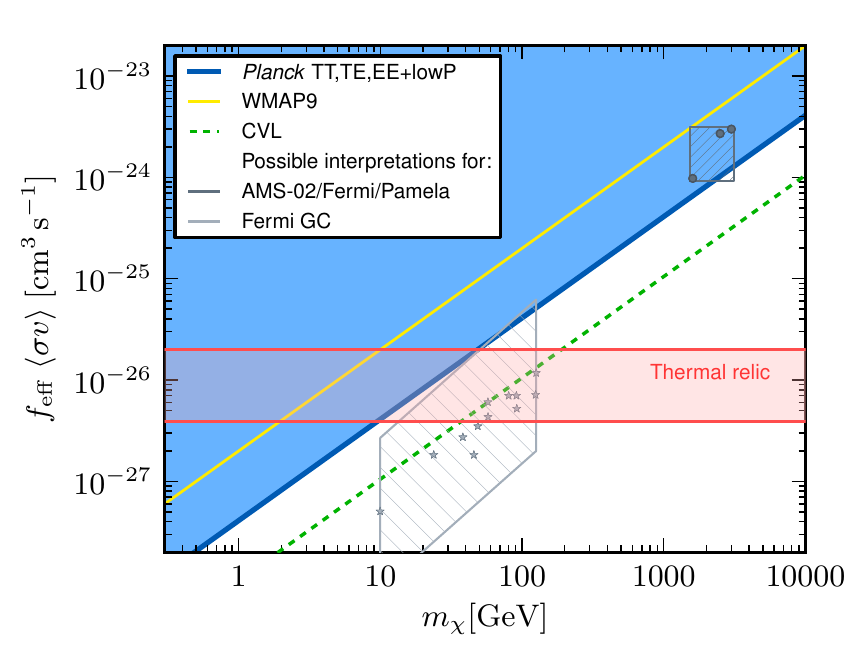}}
\caption{\label{fig:planck}
Limits on the parameter $f_\text{eff} \langle \sigma v \rangle$, as derived by the \emph{Planck} Collaboration. The blue region is ruled out by \emph{Planck} data; the region above the yellow line was previously excluded by WMAP9. The green line indicates the potential reach of a cosmic-variance-limited experiment. Reproduced from Ref.~\cite{PlanckCollaboration2015}; see that work for further details.}
\end{figure*}

\begin{figure*}
\includegraphics[width=0.5\textwidth]{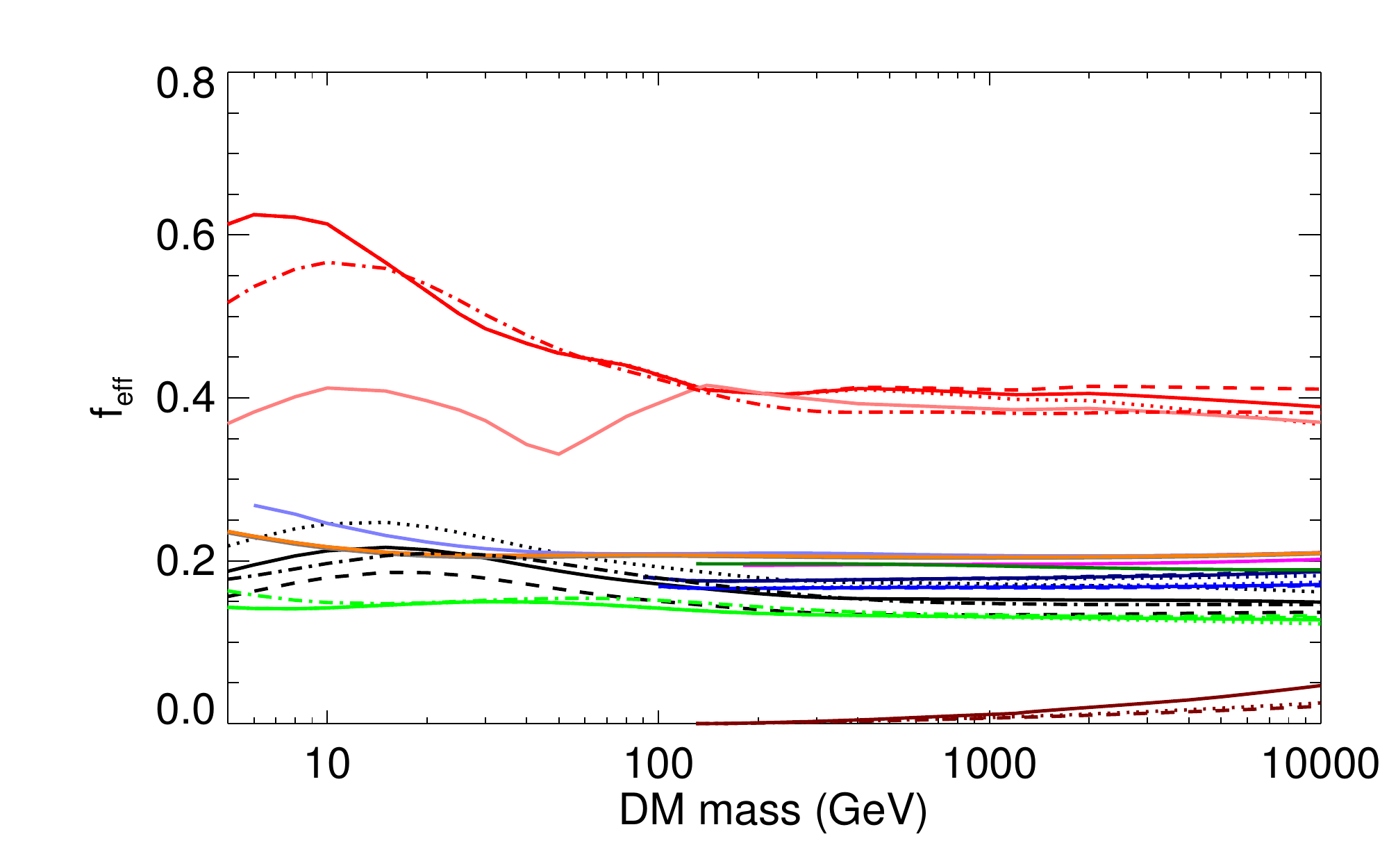}
\includegraphics[width=0.40\textwidth]{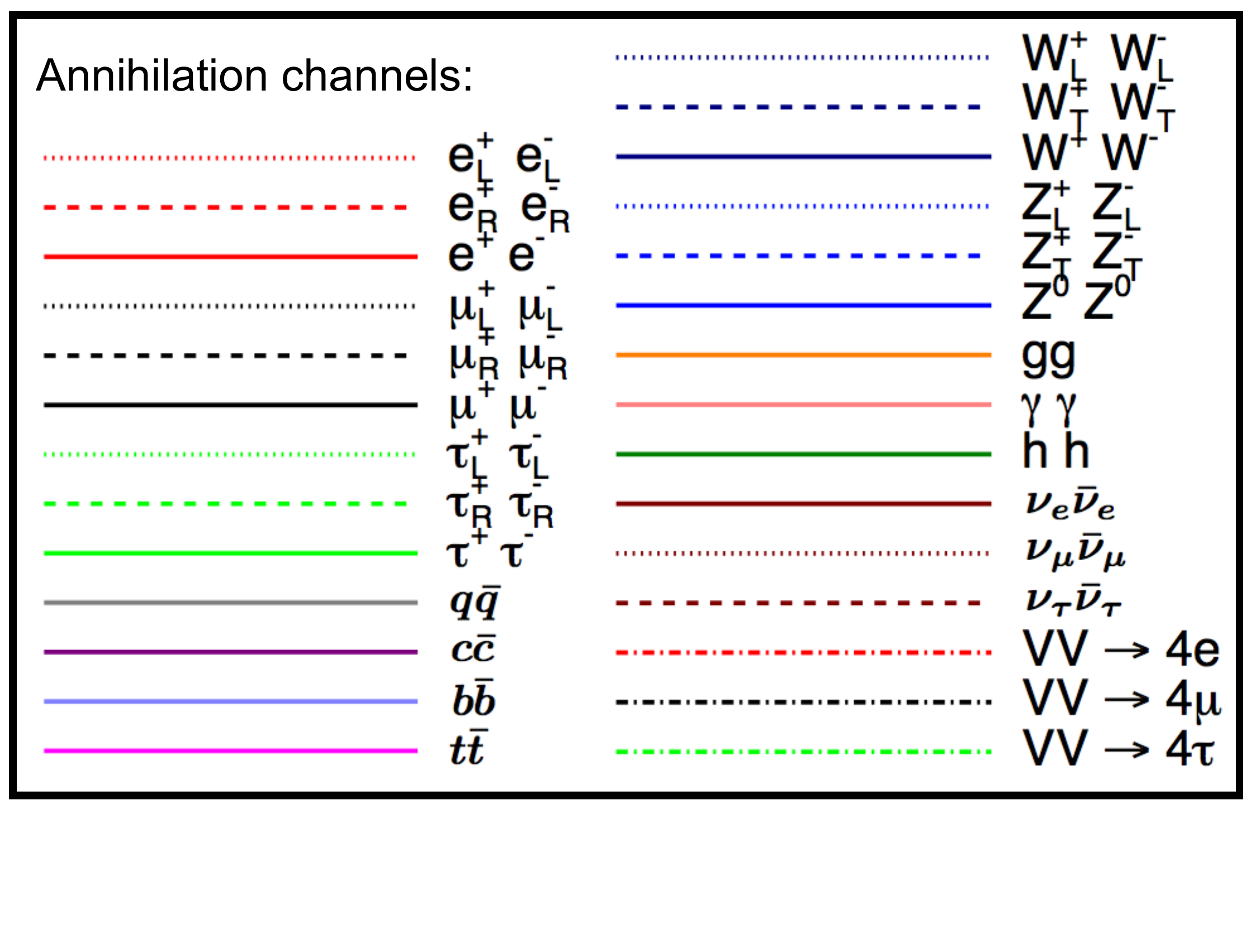}  \\
\includegraphics[width=0.45\textwidth]{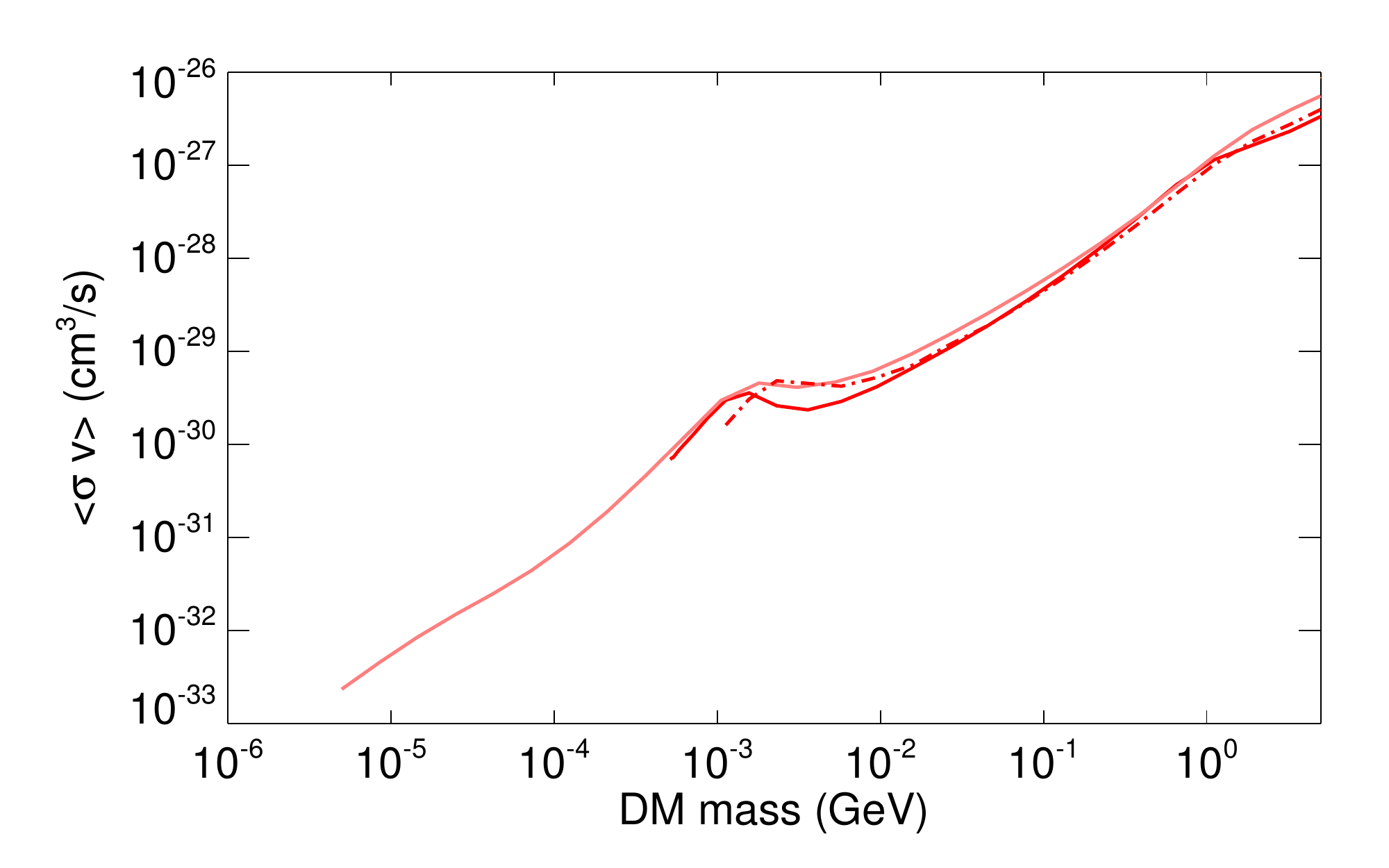}
\includegraphics[width=0.45\textwidth]{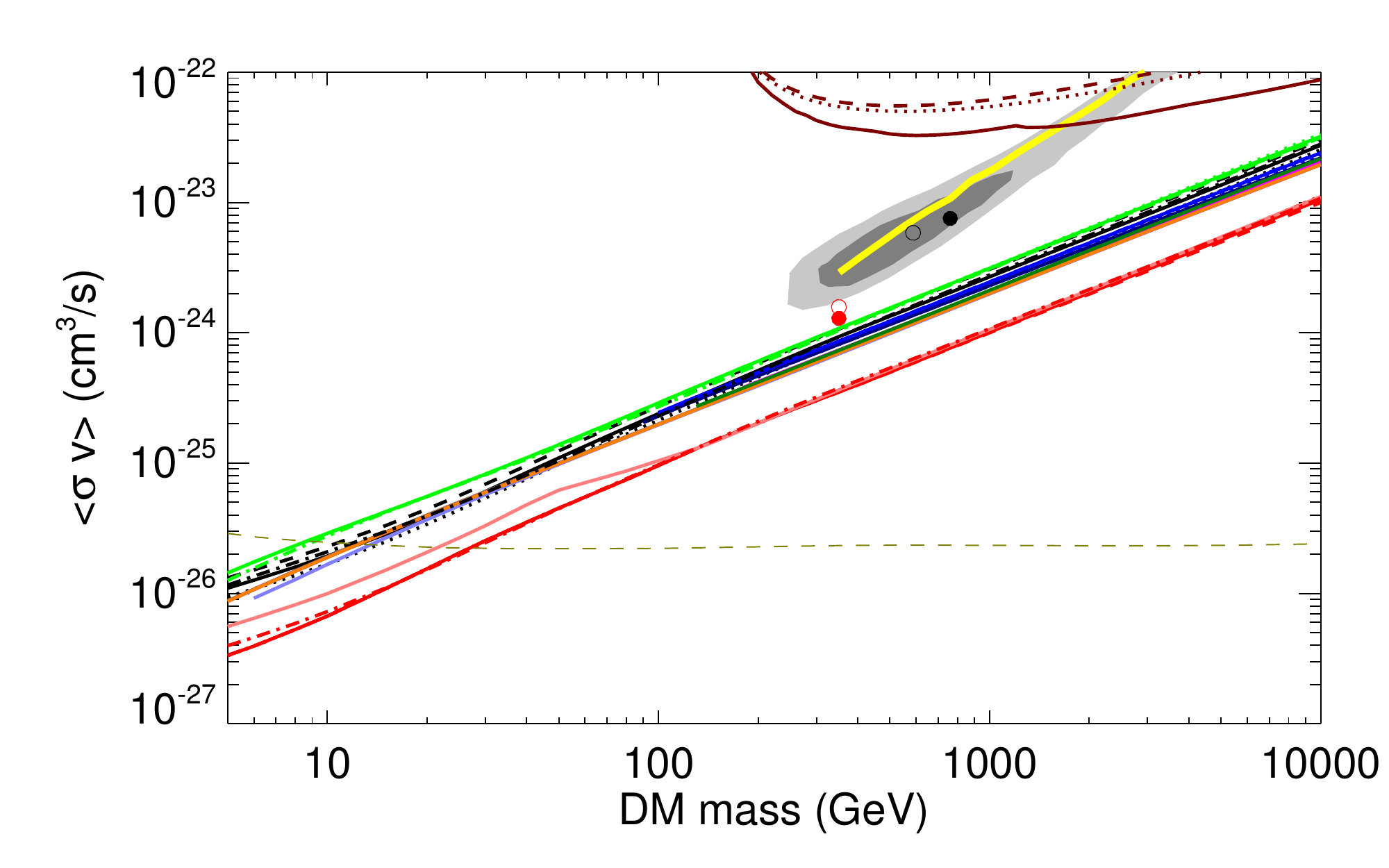}
\caption{\label{fig:annlimits}
The upper panel shows the $f_\mathrm{eff}$ coefficients as a function of DM mass for each of a range of SM final states, as indicated in the legend. The $VV \rightarrow 4X$ states correspond to DM annihilating to a pair of new neutral vector bosons $V$, which each subsequently decay into $e^+ e^-$, $\mu^+ \mu^-$ or $\tau^+ \tau^-$ (labeled by $X$). The lower panels show the resulting estimated constraints from recent \emph{Planck} results \cite{PlanckCollaboration2015} (reproduced in Fig.~\ref{fig:planck}), as a function of DM mass, for each of the channels. The left panel covers the range from keV-scale masses up to 5 GeV, and only contains results for the $e^+ e^-$, $\gamma \gamma$ and $VV \rightarrow 4e$ channels; the right panel covers the range from 5 GeV up to 10 TeV, and covers all channels provided in the \texttt{PPPC4DMID} package \cite{Cirelli:2010xx}. The light and dark gray regions in the lower right panel correspond to the $5\sigma$ and $3\sigma$ regions in which the observed positron fraction can be explained by DM annihilation to $\mu^+ \mu^-$, for a cored DM density profile (necessary to evade $\gamma$-ray constraints), taken from Ref.~\cite{Cirelli:2008pk}. The solid yellow line corresponds to the preferred cross section for the best fit 4-lepton final states identified by Ref.~\cite{Boudaud:2014dta}, who argued that models in this category can still explain the positron fraction without conflicts with non-observation in other channels. The red and black circles correspond to models with $4e$ (red) and $4\mu$ (black) final states, fitted to the positron fraction in Ref.~\cite{Lopez:2015uma}; as in that work, filled and open circles correspond to different cosmic-ray propagation models. The near-horizontal dashed gold line corresponds to the thermal relic annihilation cross section \cite{Steigman:2012nb}. Reproduced from Ref.~\cite{Slatyer2015a}.}
\end{figure*}

For the case of annihilation, it is conventional to normalize $f_\text{eff}$ to the case of a reference model where 100\% of the injected power is promptly absorbed by the gas, and roughly 1/3 of this power goes into ionization if the background ionization level is low. Choosing $f_\text{eff}=1$ for this reference model, CMB data can be used to set a limit on $f_\text{eff} \langle \sigma v_\text{rel} \rangle/m_\text{DM}$. Fig.~\ref{fig:planck} shows this limit derived from Planck data \cite{PlanckCollaboration2015}. Fig.~\ref{fig:annlimits} shows the resulting upper bounds on $\langle \sigma v_\text{rel} \rangle$ as a function of $m_\text{DM}$, for various 2-body SM final states, using the curves shown in Fig.~\ref{fig:feff} to calculate the final-state-dependent and mass-dependent $f_\text{eff}$ factors. Similar calculations can be performed for the case of decaying DM; see Ref.~\cite{Slatyer:2016qyl}.

Because the CMB constraints measure total injected power, and the effect on the CMB anisotropy spectrum is essentially model-independent up to the overall normalization factor, these limits can be applied to a very wide range of DM models. In particular, they are often the strongest available constraints for DM masses and annihilation channels where the annihilation products are difficult to detect directly with current telescopes (e.g. because low-energy electrons and positrons are deflected by the solar wind, or low-energy photons are absorbed on their way to Earth, or we have no current telescopes observing the relevant energy range, or the astrophysical backgrounds are large and difficult to characterize).

However, when the annihilation/decay products \emph{can} be observed directly, the resulting constraints are typically much stronger. We will discuss some such constraints next.

\subsection{WIMP annihilation limits from gamma rays}

The most stringent robust bounds for weak-scale DM annihilating to photon-rich channels come from observations of the Milky Way's dwarf galaxies by \emph{Fermi} \cite{Ackermann:2015zua}. These limits are publicly available as likelihood functions for the flux in each energy bin, allowing constraints to be set on arbitrary spectra\footnote{\texttt{https://www-glast.stanford.edu/pub\_data/}}; for example, for annihilation to $b$ quarks, DM masses below $\sim 100$ GeV are excluded. The VERITAS and MAGIC telescopes also set constraints on these channels from similar dwarf studies \cite{Zitzer:2015eqa,Ahnen:2016qkx}, which can dominate those from \emph{Fermi} for DM masses well above 1 TeV. Stronger high-energy limits have been presented using data from the H.E.S.S telescope \cite{Abdallah:2016ygi}, although these rely on studies of the region around the Galactic Center, and are thus more sensitive to uncertainties in the DM density profile. Observations of dwarf galaxies in gamma rays typically involve an integration over much of the volume of the dwarf, and so are less sensitive to the details of the density profile in the innermost regions; however, there can still be large uncertainties in the J-factors due to uncertainties in the DM content and distribution (e.g. Ref.~\cite{Chiappo:2016xfs}).

Note that as a general rule, air Cherenkov telescopes (ACTs), such as H.E.S.S, VERITAS and MAGIC, set the strongest limits at high energies; these telescopes are ground-based, and so can have very large effective areas. However, they lose sensitivity to gamma rays below the $\mathcal{O}(100)$ GeV energy scale; in this regime, space-based telescopes such as \emph{Fermi} play a crucial role, despite their much smaller collecting areas. \emph{Fermi} also has the advantage of being a full-sky telescope, which facilitates blind searches for signals and studies of large-scale diffuse emission, whereas current air Cherenkov telescopes have comparatively small fields of view.

\subsection{WIMP annihilation limits from cosmic rays}

\begin{figure*}
\centerline{
\includegraphics[scale=0.4]{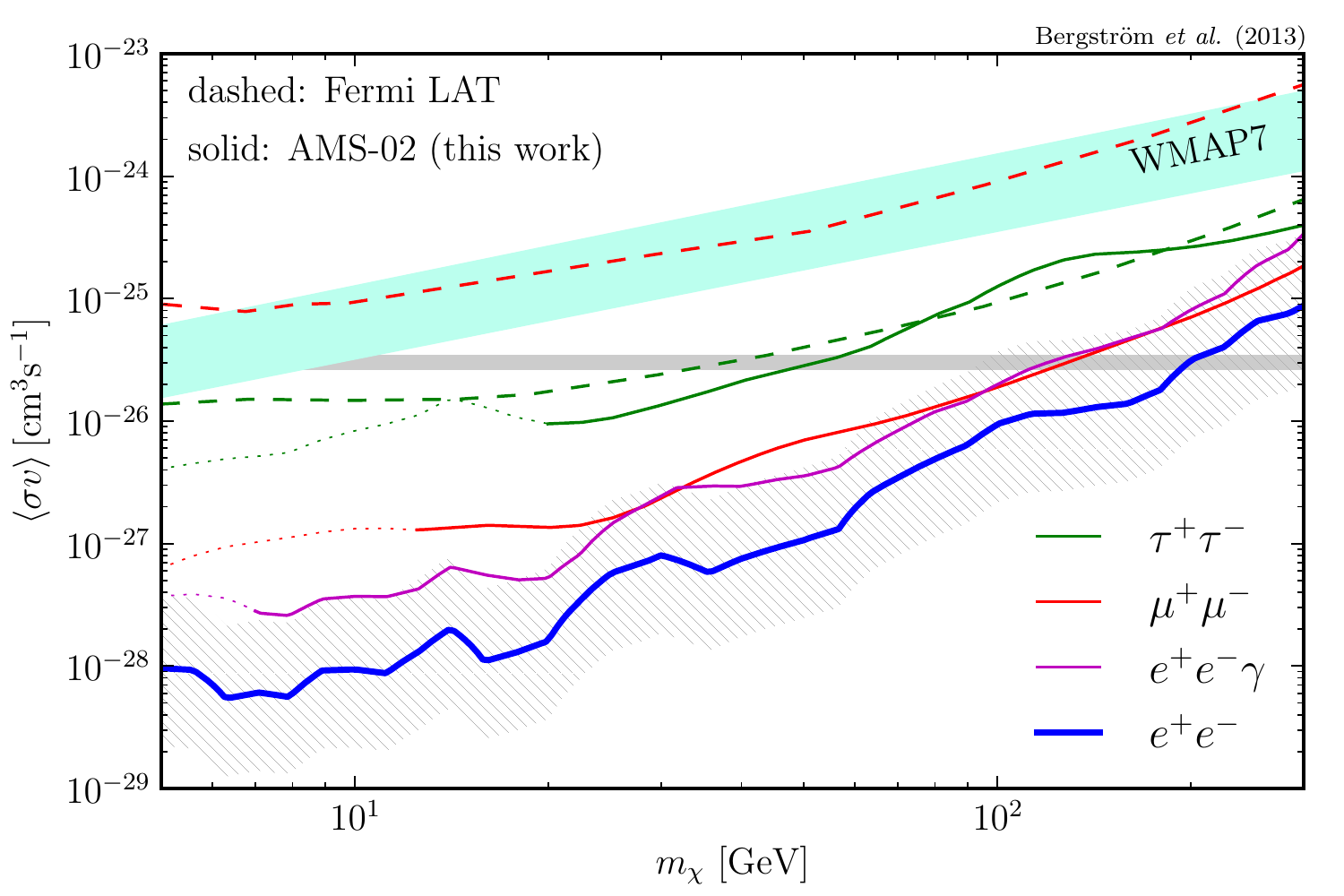}
\includegraphics[scale=0.4]{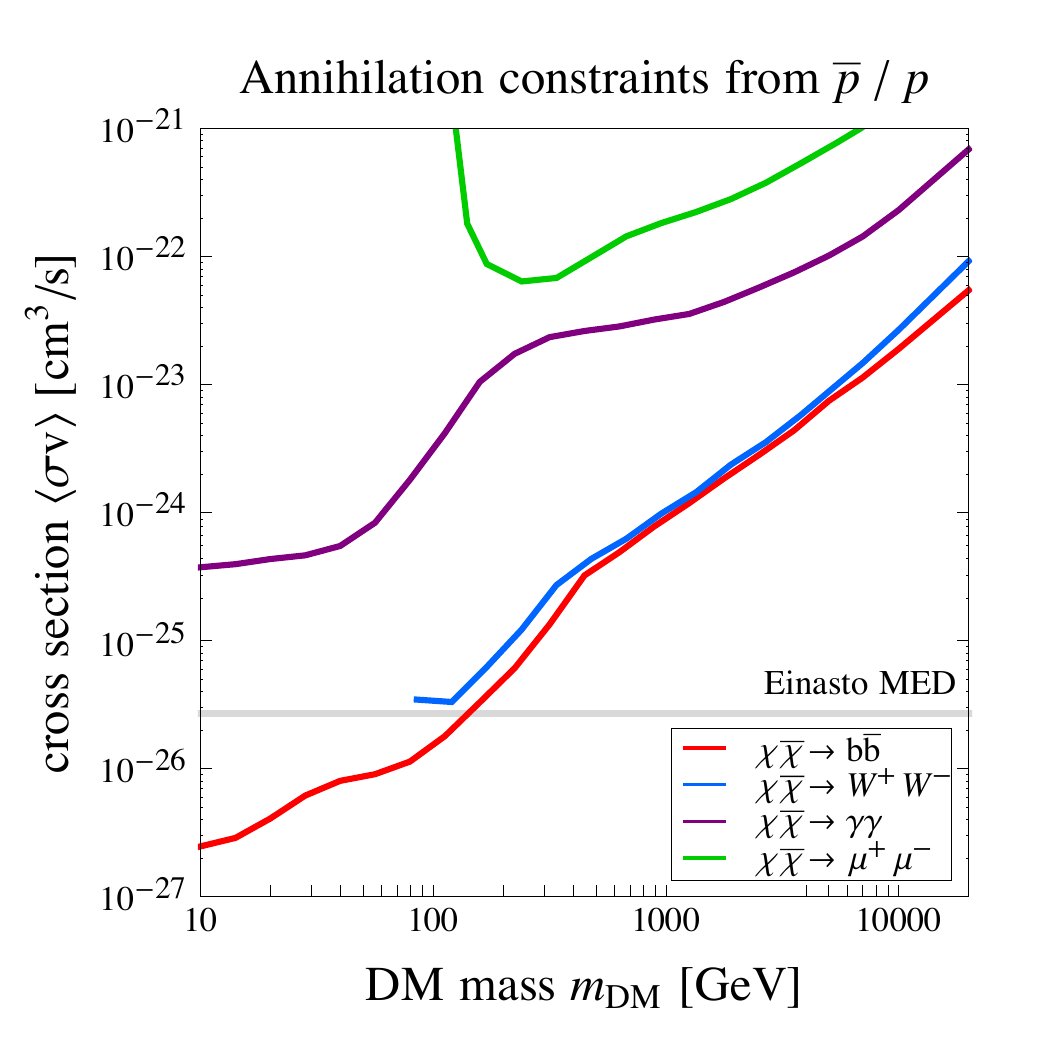}}
\caption{Limits from AMS-02 data on DM annihilation to (left) leptonic channels and (right) channels producing antiprotons. In the left panel, dashed lines indicate bounds from \emph{Fermi} observations of dwarf galaxies; the dotted
portions of the solid constraint curves from AMS-02 are potentially affected by solar modulation effects. The hatched band around the $e^+ e^-$ constraint line indicates the estimated uncertainty due to systematic uncertainties in the local DM density and energy loss rate. Reproduced from Ref.~\cite{Bergstrom:2013jra} (left panel) and Ref.~\cite{Giesen:2015ufa} (right panel).}
\label{fig:ams}
\end{figure*}

The AMS-02 instrument has presented measurements of the spectrum of a wide range of cosmic ray species, at the location of the Earth.   For DM searches the most relevant channels are positrons \cite{Bergstrom:2013jra} and antiprotons \cite{Giesen:2015ufa}, although measurements of other cosmic rays help constrain the propagation parameters discussed above. (Very recent studies, occurring after these lectures were first presented, have also claimed evidence for a potential DM signal in the antiproton data \cite{Cuoco:2016eej,Cui:2016ppb}.)

Fig.~\ref{fig:ams} displays limits on DM annihilation from AMS-02 measurements of positrons and antiprotons, which provide sensitive probes -- potentially more sensitive than the dwarf searches -- of leptonic and hadronic annihilation channels respectively. However, these constraints are subject to substantial systematic uncertainties, associated with cosmic-ray propagation, the effects of the Sun's magnetic field, and (in the hadronic case) the production cross section for antiprotons.

\subsection{Line limits from the Galactic Center}

\begin{figure*}
\centerline{
\includegraphics[width=0.6\textwidth]{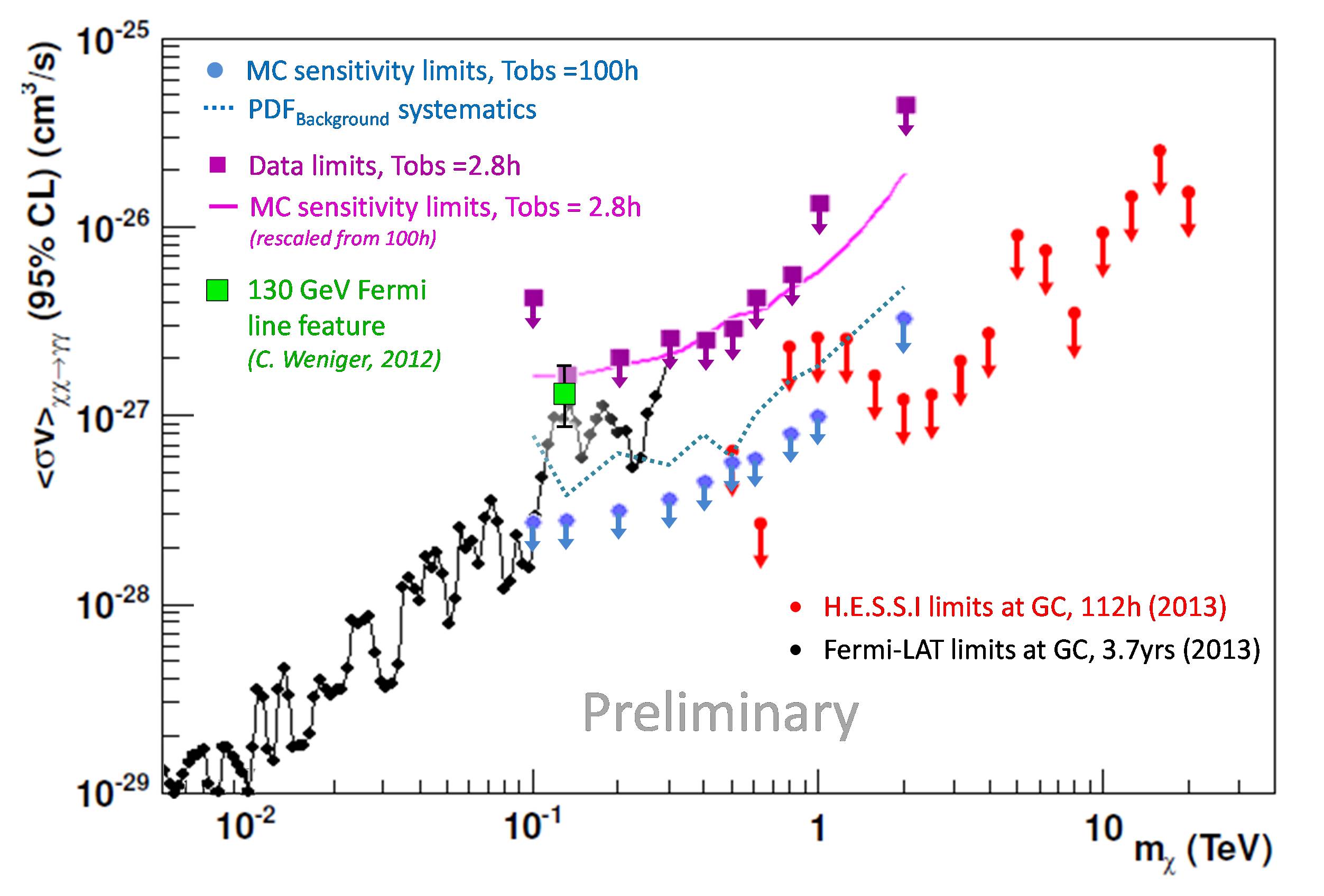}}
\caption{Upper bounds on the cross section for DM annihilation to $\gamma \gamma$, from \emph{Fermi} (black points) and H.E.S.S (red and purple points). Blue points indicate the forecast sensitivity for a future 100-hour observation with H.E.S.S. Limits are at the 95\% confidence level. Reproduced from Ref.~\cite{Kieffer:2015nsa}.}
\label{fig:lines}
\end{figure*}

For gamma-ray lines, as discussed above, astrophysical backgrounds are low. Thus the imperative is to optimize statistics, and it makes sense to look toward the Galactic Center. H.E.S.S \cite{Abramowski:2013ax} and \emph{Fermi}  \cite{Albert2014} have presented limits on the possible gamma-ray line strength, as summarized in Fig.~\ref{fig:lines}. 

\begin{table}
	\begin{center}
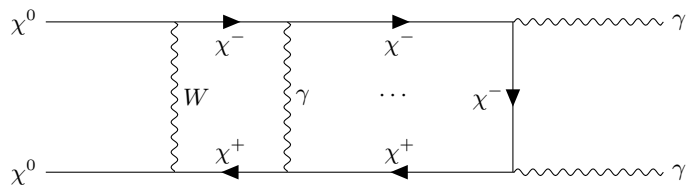
	
	\begin{tikzpicture}
		  \begin{feynman}
			\vertex (a1) {\(\chi^0\)}; 
			\vertex[below=2cm of a1] (a2) {\(\chi^0\)};
			\vertex[right=2cm of a1] (b1);
			\vertex[right=2cm of a2] (b2);
			\vertex[right=1.5cm of b1] (c1);
			\vertex[right=1.5cm of b2] (c2);
			\vertex[right=3cm of c1] (d1);
			\vertex[right=3cm of c2] (d2);
			\vertex[right=2cm of d1] (e1) {\(\gamma\)};
			\vertex[right=2cm of d2] (e2) {\(\gamma\)};
			\vertex[below right =0.8cm and 1.1cm of c1] (dots) {\(\cdots\)};
			\diagram*{
				(c2) -- [fermion, edge label'=\(\chi^+\)] (b2);
				(b1) -- [fermion, edge label'=\(\chi^-\)] (c1);
				(c1) -- [fermion, edge label'=\(\chi^-\)] (d1);
				(d2) -- [fermion, edge label'=\(\chi^+\)] (c2);	
				(a1) -- (b1);
				(a2) -- (b2);
				(d2) -- [boson] (e2);
				(d1) -- [boson] (e1);
				(b2) -- [boson, edge label'=\(W\)] (b1);	
				(c2) -- [boson, edge label'=\(\gamma\)] (c1);	
				(d1) -- [fermion, edge label'=\(\chi^-\)] (d2);	
			};
		  \end{feynman}
		\end{tikzpicture}
	\end{center}
	\captionof{figure}{Example of annihilation of a supersymmetric wino $\chi^0$ to photons, via a long-range potential mediated by exchange of weak gauge bosons.}
	\label{fig:winodiagram}
\end{table}

\begin{figure*}[t!]
\centerline{
\includegraphics[scale=0.46]{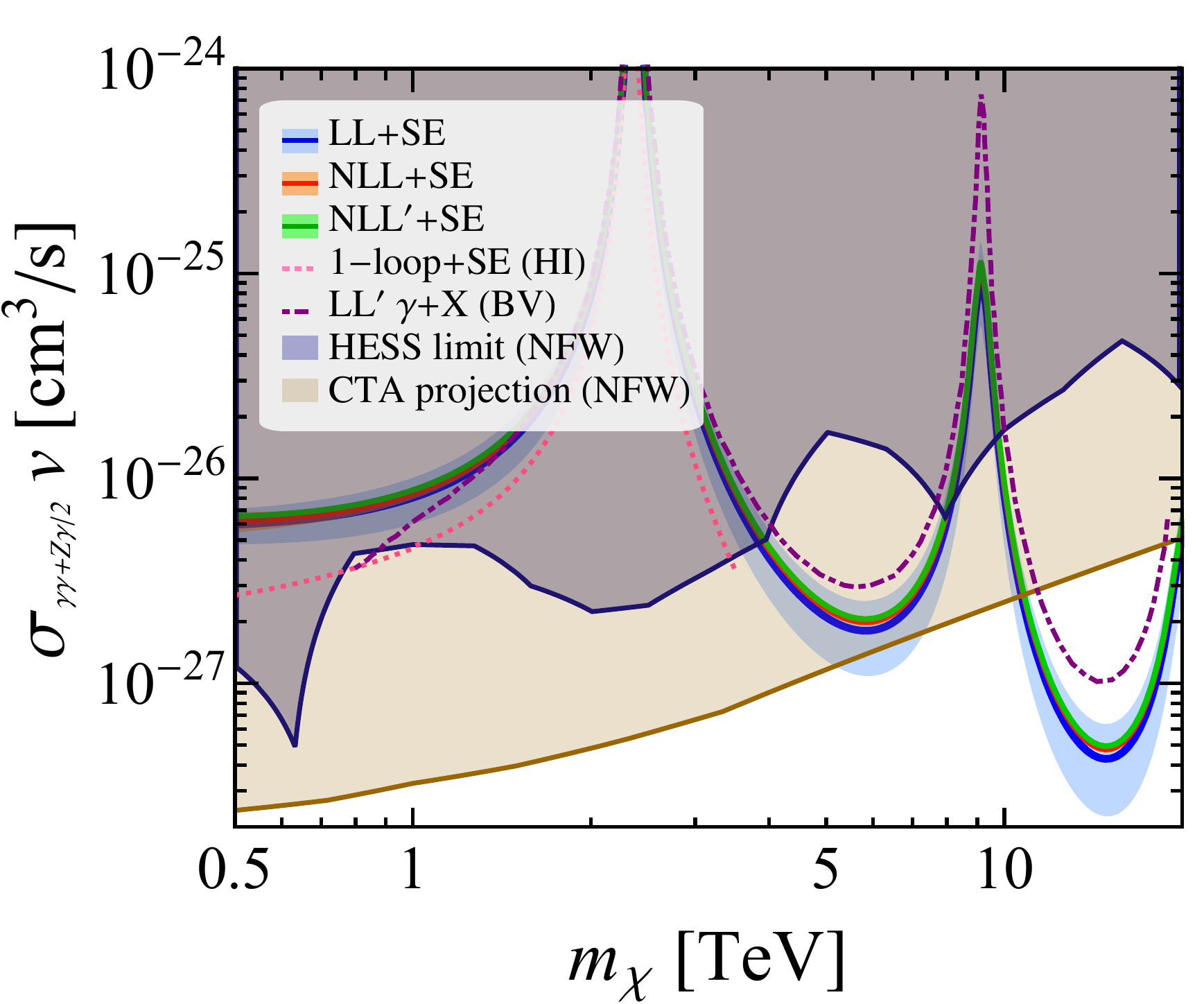}}
\caption{The cross section for wino annihilation to produce a photon line, including the Sommerfeld enhancement and resummed logarithmic corrections, compared to current bounds from H.E.S.S. and projected ones from CTA, in the latter case assuming 5 hours of observation time. The green band indicates the most accurate and up-to-date calculation, with the pink dotted line and red and blue bands indicating earlier results. Also shown (purple dot-dashed line) is the rate for the semi-inclusive process $\gamma + X$ calculated to LL$^{\prime}$ in Ref.~\cite{Baumgart:2015bpa}. Reproduced from Ref.~\cite{Ovanesyan:2016vkk}.}
\label{fig:winolimit}
\end{figure*}

Note that while the usual expectation is that the line cross section will be well below the thermal relic value, there are caveats to this statement; in particular, if there are charged particles in the spectrum of the theory, close in mass to the DM, then the line cross section can be unexpectedly large. This is particularly true in cases where a long-range potential couples two-particle DM states to two-particle states involving the charged particles -- which is the case, for example, for pure wino DM in supersymmetric models. When $m_\text{DM} > m_W/\alpha_W$, the exchange of weak gauge bosons becomes effectively a long-range force, associated with a Sommerfeld enhancement that can readily be 1-2 orders of magnitude; the line cross section can be enhanced even further, since the long-range W-exchange potential allows any pair of winos to effectively oscillate into a chargino-chargino state, which can annihilate to $\gamma \gamma$ at tree level (see Fig.~\ref{fig:winodiagram}). Fig.~\ref{fig:winolimit} shows how the prediction for the line cross section for wino DM compares to current and future constraints, assuming the pure wino constitutes 100\% of the DM (this requires non-thermal production for masses not in the 2-3 TeV range). This is an example of an indirect search probing regions of high-mass DM parameter space that cannot be explored by colliders.

\begin{figure*}
\centerline{
\includegraphics[scale=0.8]{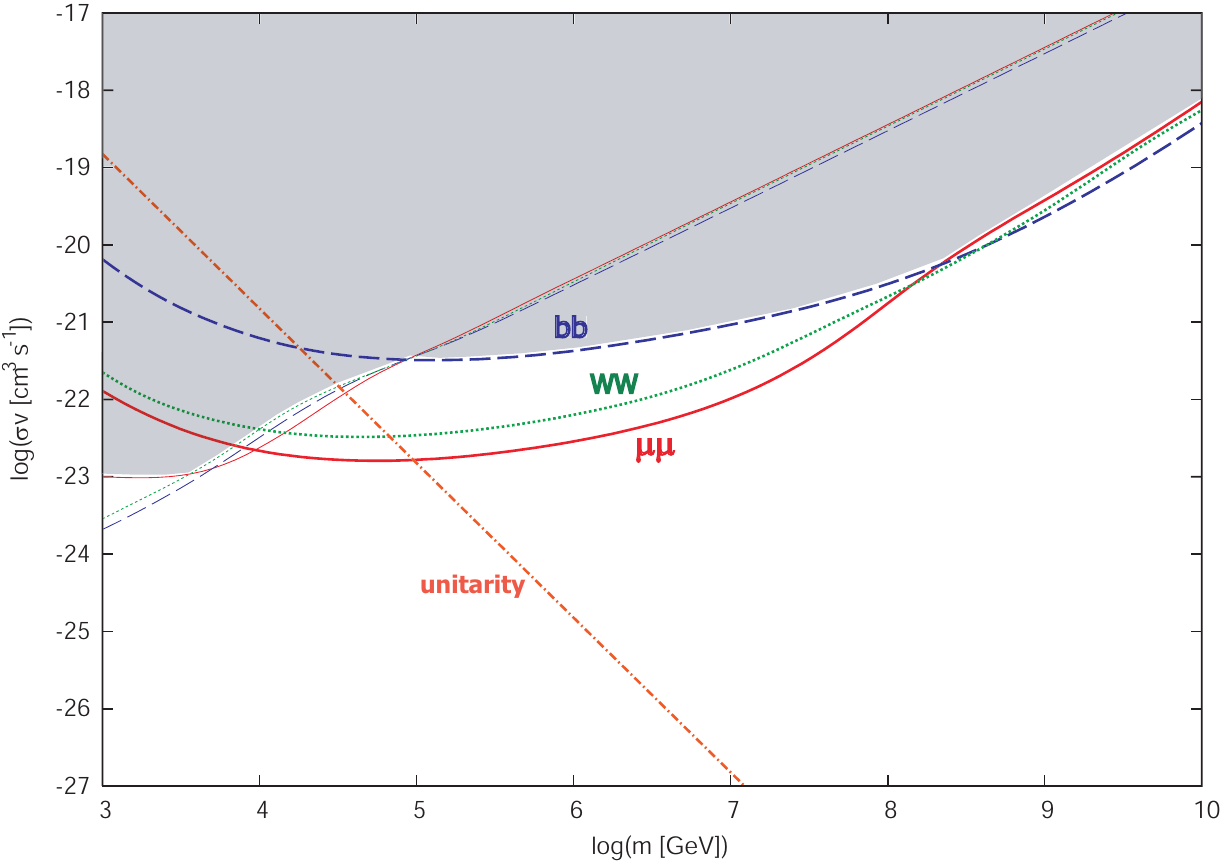}}
\caption{Neutrino (thick curves) and cascade gamma-ray (thin curves) constraints on the annihilation
cross section of very heavy DM. The shaded region is determined by taking the more stringent of the neutrino and gamma-ray bounds, for the least constrained of the three channels. Reproduced from Ref.~\cite{Murase:2012xs}.}
\label{fig:highmass}
\end{figure*}

\subsection{Annihilation of very heavy DM}

For DM well above the TeV scale, constraints can be set using either gamma-ray telescopes (such as H.E.S.S, VERITAS, MAGIC and \emph{Fermi}) or neutrino telescopes such as IceCube. Fig.~\ref{fig:highmass} shows current limits; here the modeling of the signal includes contributions from DM substructure, and modeling of energy losses for gamma rays traveling intergalactic distances. Sufficiently high-energy photons can pair produce on the interstellar radiation field, producing an electron-photon cascade that results in a spectrum of gamma rays at lower energies; thus often observations from \emph{Fermi}, which can observe photons in the 1-100 GeV range, are actually more constraining than from experiments that only observe higher-energy gamma rays. Note that the requirement of unitarity strongly constrains the annihilation rate at sufficiently high mass scales.

\subsection{Heavy dark matter decays}

\begin{figure*}
\centerline{
\includegraphics[width=0.8\textwidth]{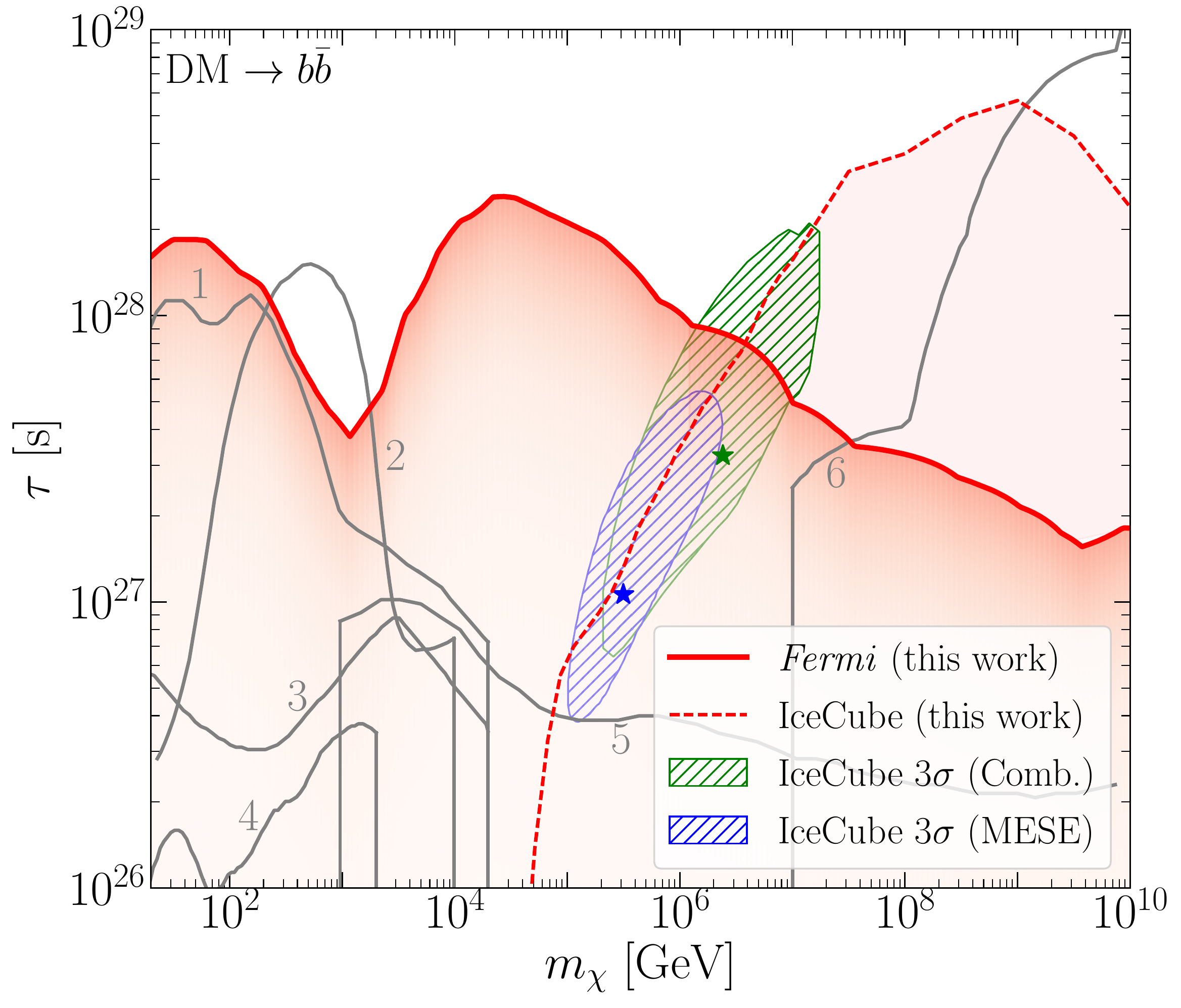}}
\caption{Lower bounds on the decay lifetime for DM decaying to $b$ quarks. The red line is determined by Ref.~\cite{Cohen:2016uyg} using \emph{Fermi} data; gray lines with numbers denote existing bounds using data from \emph{Fermi} (2,3,5), AMS-02 (1,4), and PAO/KASCADE/CASAMIA (6). The hashed green (blue) region suggests parameter space where DM decay may provide a $\sim3 \sigma$ improvement to the description of the combined maximum likelihood IceCube neutrino flux. The red dotted line provides a limit if a combination of
DM decay and astrophysical sources are responsible for the observed high-energy neutrinos. Reproduced from Ref.~\cite{Cohen:2016uyg}.}
\label{fig:decay}
\end{figure*}

As for annihilation, heavy decaying DM can be constrained by observations from gamma-ray and neutrino telescopes. Fig.~\ref{fig:decay} summarizes the results of several analyses for the example $\bar{b} b$ channel; we see that generically lifetimes shorter than $\sim 10^{27-28}$ s can be ruled out, across a very large mass range.

\subsection{Light dark matter decays}

Decays of light DM, below the GeV scale, cannot be easily constrained by \emph{Fermi}, for which effective area and angular resolution degrade rapidly at energies below a GeV. DM below the $\sim 100$ MeV scale also cannot decay into hadronic channels, which usually suppresses photon signals (due to lack of $\pi^0$ production), unless the DM decays directly to photons (which typically has a small branching ratio since the DM is uncharged).

\begin{figure}
  \centerline{  \includegraphics[width=0.5\textwidth]{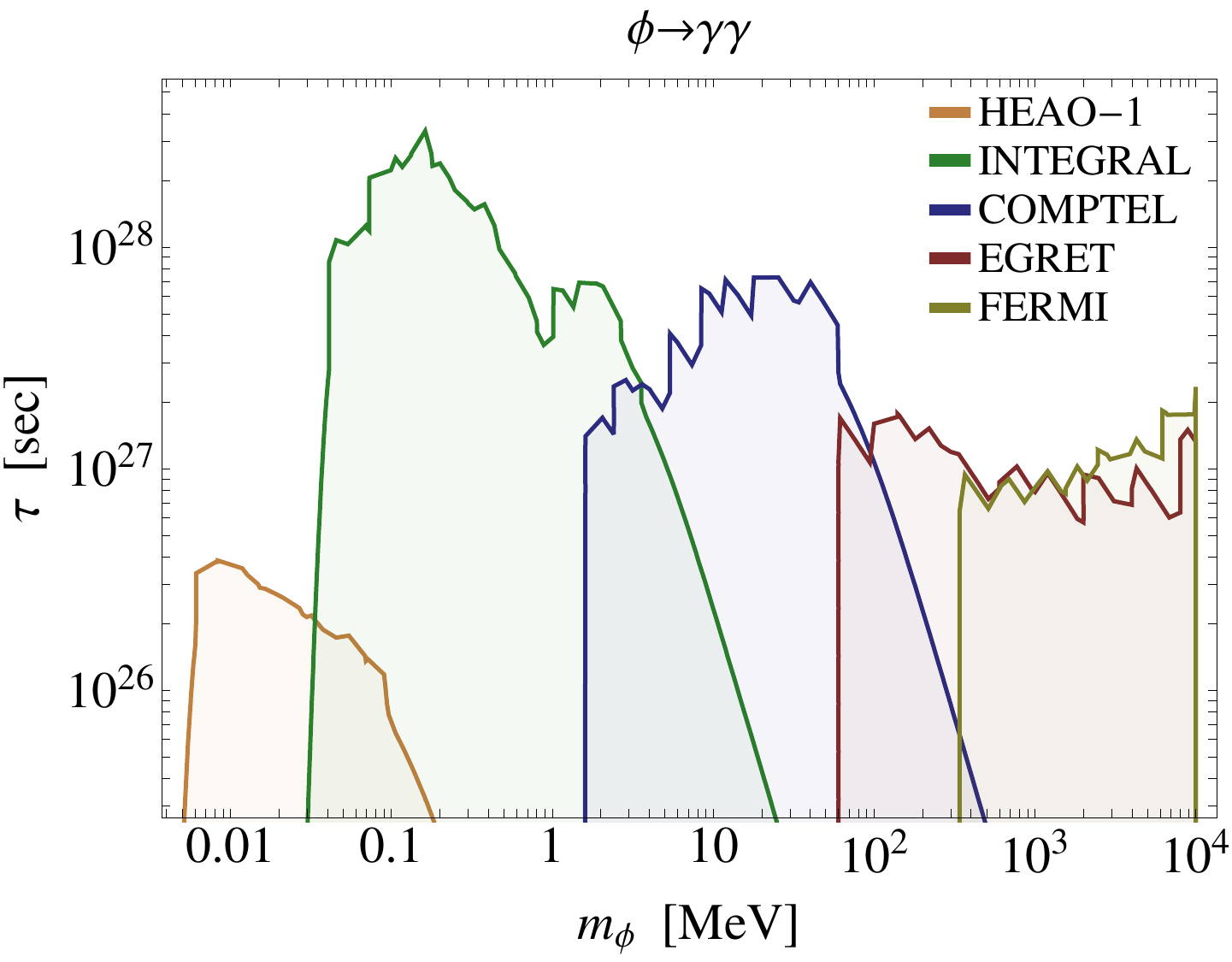} 
  \includegraphics[width=0.45\textwidth]{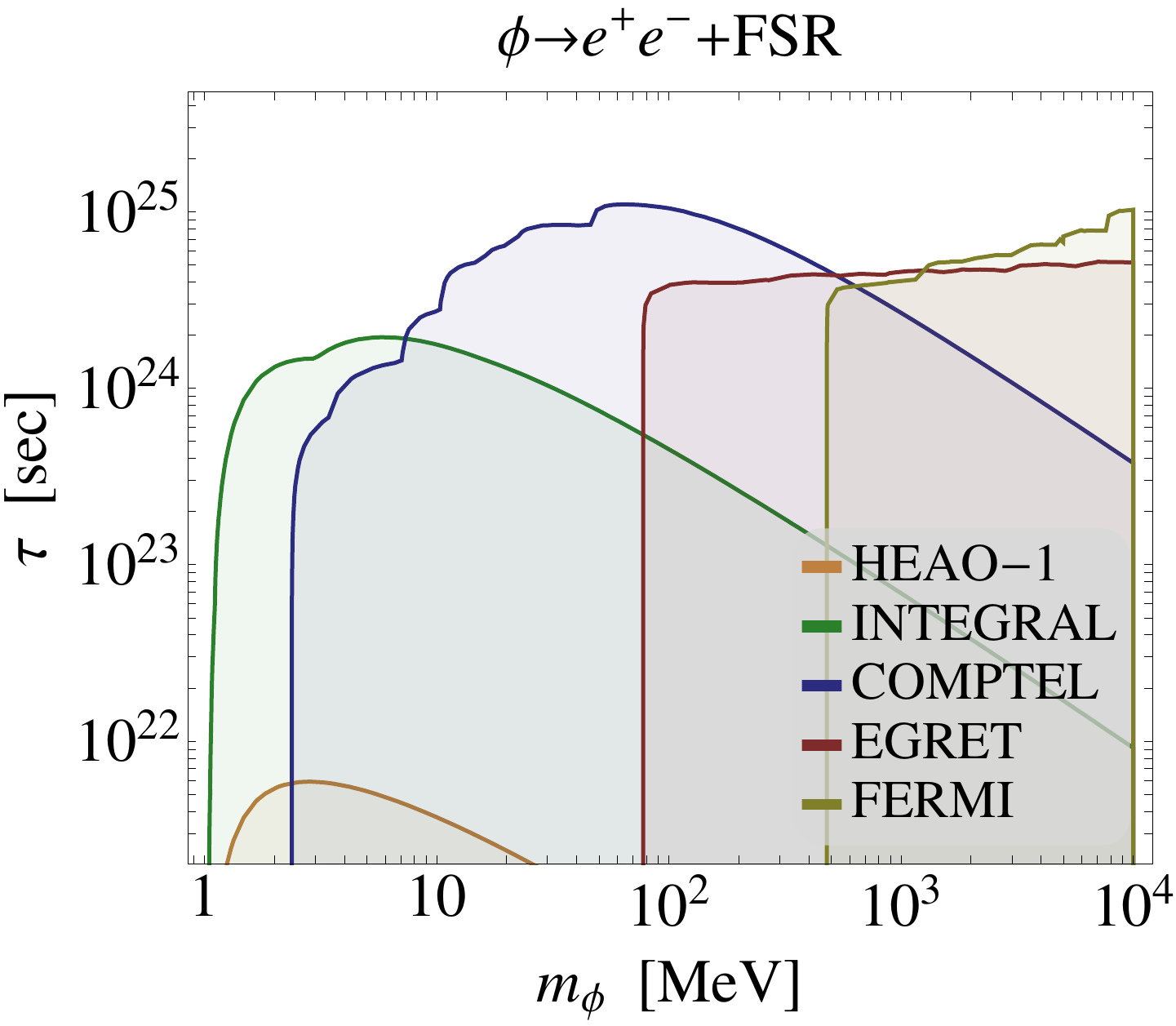} }  
  \caption{
   Lower bounds on the DM decay lifetime, for decay to $\gamma \gamma$ (left panel), or decay to  $e^+e^-+\gamma$ (right panel), from present-day diffuse photon searches with a range of experiments. Reproduced from Ref.~\cite{Essig2013}.}
\label{fig:lightdec}
\end{figure}

\begin{figure}
  \centerline{  \includegraphics[width=7cm]{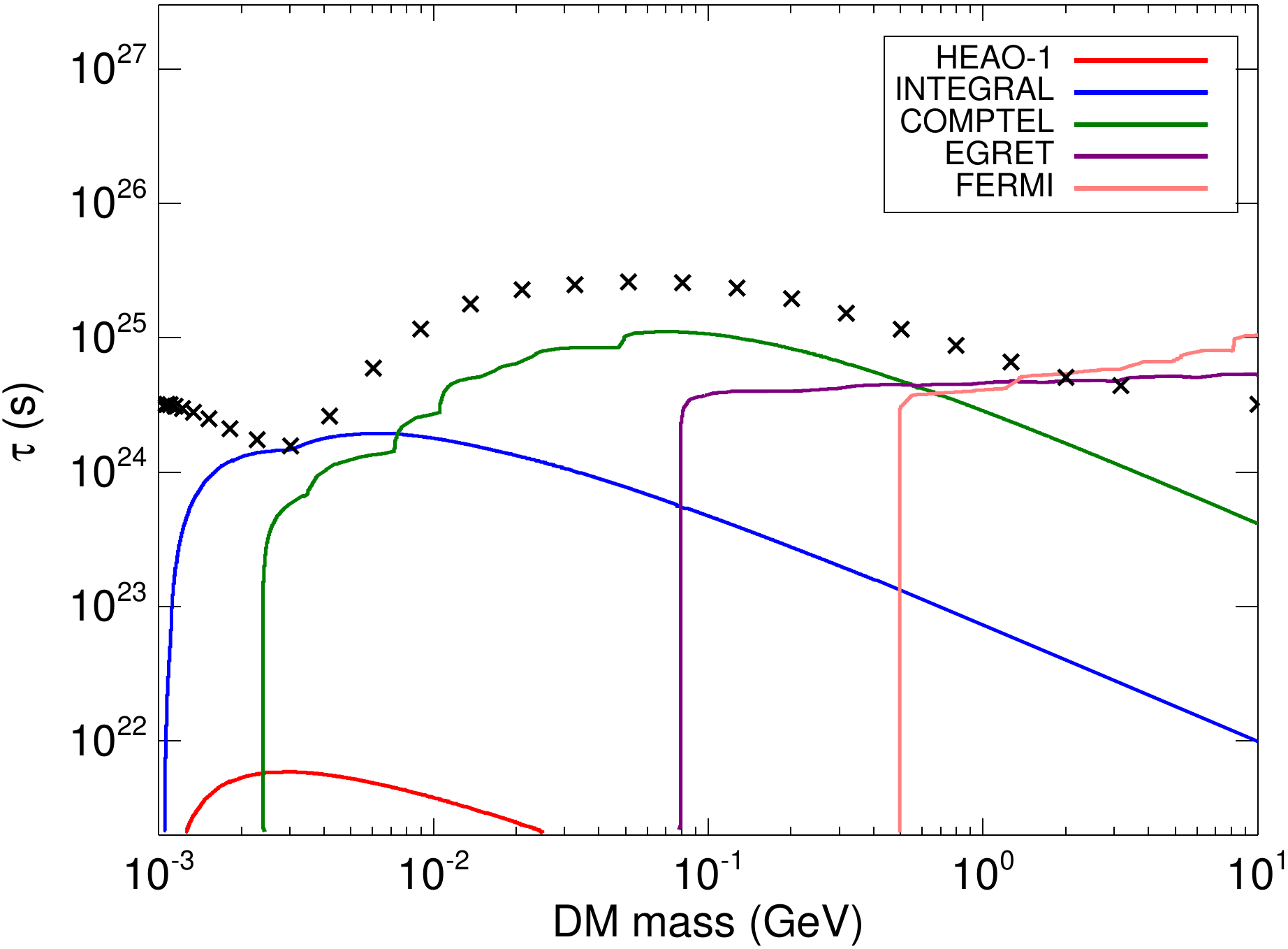} }  
  \caption{
   Lower bounds on the DM decay lifetime, for decay to $e^+e^-$, from present-day diffuse photon searches (colored lines) and from the cosmic microwave background (black crosses). Reproduced from Ref.~\cite{Slatyer:2016qyl}.}
\label{fig:compare}
\end{figure}

Nonetheless, lower-energy gamma-ray and hard X-ray telescopes do exist, and have set limits on the DM decay rate to final states involving line photons, and 3-body final states where a photon is produced by internal bremsstrahlung or final state radiation. The limits shown in Fig.~\ref{fig:lightdec} arise from studies of the diffuse photon background at these energies \cite{Essig2013}. We see that these searches are sensitive to lifetimes around $10^{26-28}$ s if line photons are produced, and to lifetimes in the neighborhood of $10^{24-25}$ s in the case of $e^+ e^-$ final states. For the case of decay to electrons and positrons with photons only produced by FSR, the limits from the CMB discussed earlier are competitive, as shown in Fig.~\ref{fig:compare}.

In addition to these limits, there are a few channels in which possible DM signals have been detected. I will now summarize the status of a few of these.

\subsection{The cosmic ray positron excess}

By the arguments given in section \ref{sec:crs}, the ratio of secondary to primary cosmic rays is generally expected to decrease with increasing energy. However, a rise in the positron fraction -- the ratio of positron flux to total electron+positron flux -- at energies above $\sim 10$ GeV was observed by the PAMELA experiment in 2008 \cite{Adriani:2008zr}, and has since been confirmed by \emph{Fermi} \cite{FermiLAT:2011ab} and AMS-02 \cite{Aguilar:2013qda}. The AMS-02 measurement has much smaller uncertainties than the original PAMELA detection, and extends to higher energies; the positron fraction appears to continue to rise up to energies of $\sim 300$ GeV. At higher energies, there is some evidence of a turnover, but the statistical uncertainties are large.

One possible explanation for this excess is DM annihilation or decay, producing additional primary positrons. Other possibilities include positrons sourced by pulsars (e.g. Ref.~\cite{Profumo:2008ms}), secondary positrons re-accelerated in some way (e.g. inside supernova remnants) \cite{Blasi:2009hv}, or some substantial modification to our understanding of cosmic-ray production or propagation \cite{Blum:2013zsa}. Under the DM-origin hypothesis, the DM must be quite heavy, with mass at least several hundred GeV; the annihilation or decay must occur primarily into leptonic channels to avoid constraints from antiproton and gamma-ray searches; and if annihilation is responsible, the cross section must be well above the thermal relic value (e.g. Ref.~\cite{Boudaud:2014dta}). While all of these features can be found in DM models \cite{ArkaniHamed:2008qn}, the required parameters are in tension or apparently excluded by the constraints discussed above (e.g. the annihilation explanation is generically in tension with CMB bounds \cite{Slatyer2015a}, and the decay explanation with limits from gamma-ray observations \cite{Dugger:2010ys,Cohen:2016uyg}).

Anisotropy in cosmic-ray arrival directions could potentially probe the distribution of the positron sources. However, Galactic magnetic fields scramble the arrival direction, and consequently the expected anisotropy is small, below the 1\% level, even if the source is a single nearby pulsar (e.g. Ref.~\cite{Linden:2013mqa}). Current measurements by \emph{Fermi} \cite{2010PhRvD..82i2003A} and AMS-02 \cite{Aguilar:2013qda} find no evidence for anisotropy, but are not sensitive to such small anisotropies in any case. However, more sensitive anisotropy measurements could be obtained using observations of cosmic rays by atmospheric Cherenkov telescopes \cite{Linden:2013mqa}; while designed to observe high-energy gamma-rays, these telescopes are sensitive to cosmic-ray collisions with the atmosphere (in fact this is their main background).

\subsection{The 3.5 keV line}

An apparent spectral line at an energy of 3.5 keV was discovered in XMM-Newton observations of galaxy clusters in 2014 \cite{Bulbul:2014sua,Boyarsky:2014jta}. The significance of the signal was (at the time) roughly $4\sigma$. A recent review of this potential signal and its possible interpretations has been given by Ref.~\cite{Abazajian:2017tcc}; here I will summarize some key points.

A large number of follow-up observational studies have been performed; the signal has now also been detected tentatively in the Galactic Center \cite{Boyarsky:2014ska} and the cosmic X-ray background \cite{Cappelluti:2017ywp}. Observations of the Draco dwarf galaxy have yielded mixed results, with claims of both a faint signal and a strong exclusion \cite{Jeltema:2015mee,Ruchayskiy:2015onc}. However, studies of M31 with Chandra \cite{Horiuchi:2013noa}, stacked galaxies with Chandra and XMM-Newton \cite{Anderson:2014tza}, and dwarf galaxies with XMM-Newton \cite{Malyshev:2014xqa}, have not found a signal and have instead set stringent limits.

The simplest DM-related explanation is a decaying sterile neutrino with a mass around 7 keV. The sterile neutrino is a long-standing DM candidate, and if its mass is above a few keV, it can be sufficiently cold to evade constraints on warm DM (e.g. \cite{Abazajian:2017tcc} and references therein). However, the simple DM decay model is very predictive: the signal from any body should be proportional to the total amount of DM in that body. This hypothesis appears to be in some tension with the null results described above. There is also a (disputed) claim in the literature that the spatial morphologies of the signals observed from the Galactic Center and Perseus cluster are incompatible with decaying DM; see Refs.~\cite{Carlson:2014lla, Abazajian:2017tcc} for competing viewpoints.

There are several alternative DM-related possibilities that are less predictive, and hence less constrained. One example is ``exciting dark matter'' \cite{Finkbeiner:2014sja,Cline:2014vsa}; in this scenario the DM itself may be much heavier than the keV scale, but it has a metastable excited state 3.5 keV above the ground state. This state is excited by DM-DM collisions, and subsequently decays producing a 3.5 keV photon. The rate of excitation would scale as the DM density squared, with some non-trivial velocity dependence; this combination of parameters is far less constrained than the total DM content of an object, and could e.g. explain why no signal is seen in dwarfs (the typical DM velocity is too low to excite the excited state) while appearing in galaxy clusters (where the typical DM velocity is much higher). Another, independent, possibility is that the DM might decay producing a 3.5 keV axion-like particle (ALP), which could convert to a 3.5 keV photon in an external magnetic field \cite{Conlon:2014xsa}; the signal would then depend on the ambient magnetic field, leading to widely varying signals from different systems \cite{Alvarez:2014gua}.

There is an ongoing debate over possible contamination from potassium and chlorine plasma lines; a spectral line at a few keV is much easier to mimic with atomic processes than a gamma-ray line (see e.g. Refs.~\cite{Jeltema:2014qfa, Boyarsky:2014paa, Bulbul:2014ala, Jeltema:2014mla} for discussion). There are several known X-ray lines close to 3.5 keV and their strength can depend sensitively on the plasma temperature. Charge-exchange reactions can also produce 3.5 keV emission, and may give rise to some or all of the observed feature \cite{Gu:2015gqm, Shah:2016efh} (see also the discussion in Ref.~\cite{Cappelluti:2017ywp}, which attempts to exclude this explanation).

Energy resolution may be the key to distinguishing between DM and astrophysical origins for the line. An instrument with sufficiently good energy resolution could potentially resolve the putative 3.5 keV line at an energy distinct from any of the known atomic lines; with eV-scale energy resolution, it would be possible to measure the Doppler broadening due to the velocity of DM in the Galactic halo, if the signal originates from DM decay. The Hitomi telescope had the required energy resolution, but failed a few days after launch; a short observation of the Perseus cluster \cite{Aharonian:2016gzq} did not find evidence for a $\sim3.5$ keV line, in tension with earlier measurements claiming a bright line in Perseus, but the limits on the signal strength do not probe the DM decay explanation (fully explaining the claimed bright signal in Perseus is challenging in the context of DM decay).

\begin{figure}
\centerline{\includegraphics[width=8cm]{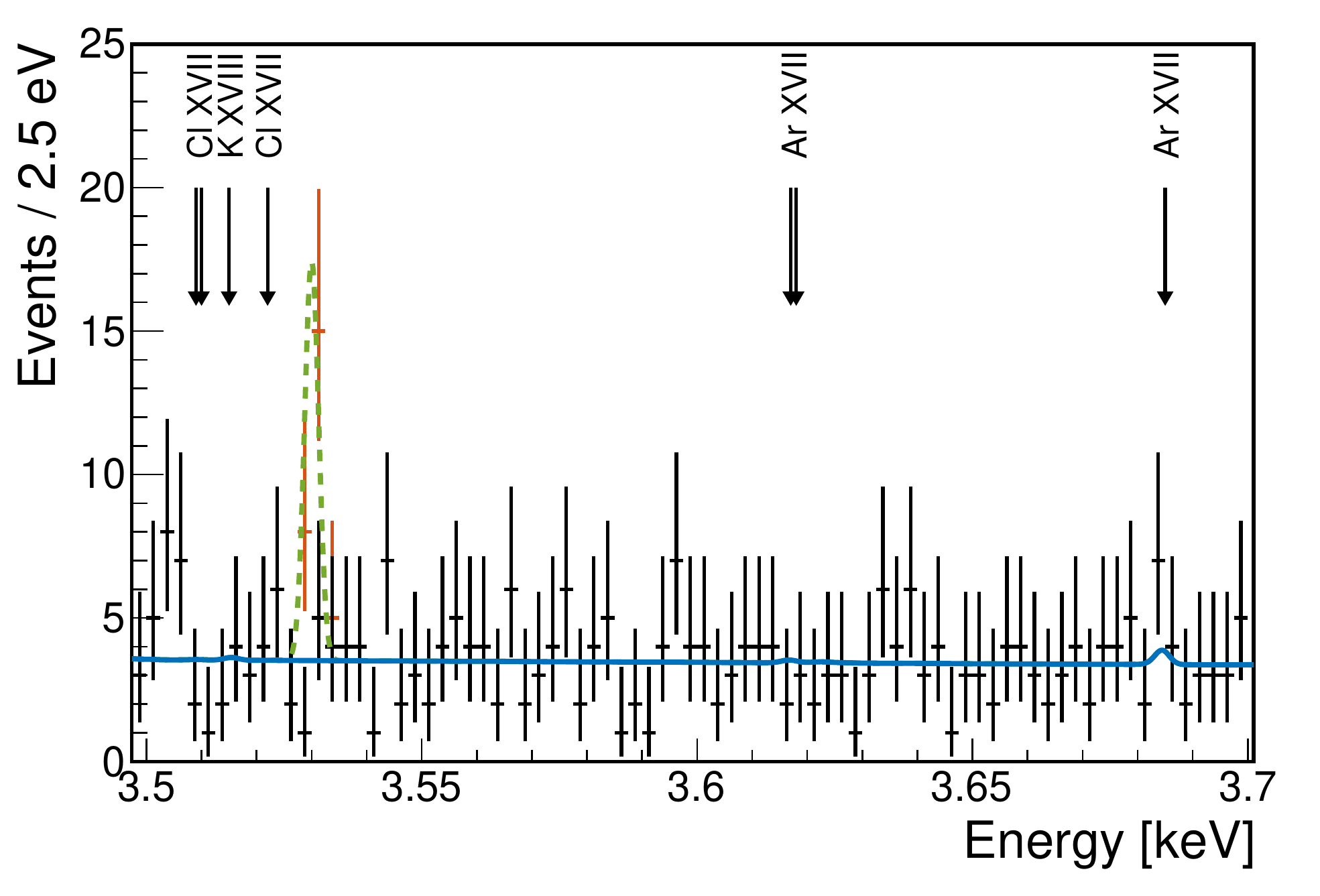}}
\caption{Micro-X mock data for the energy range of interest for the 3.5 keV line, with (red) and without (black) a signal, with background
model (blue line) and signal model (dashed green line) overlaid. Reproduced from Ref.~\cite{Figueroa-Feliciano:2015gwa}.}
\label{fig:microx}
\end{figure}

One possible mission that could probe the 3.5 keV line is the Micro-X sounding rocket program \cite{Figueroa-Feliciano:2015gwa}. By placing high-resolution X-ray spectrometers on suborbital sounding rockets, this approach would achieve excellent energy resolution -- as low as 3 eV -- for modest cost. The exposure would be short -- 5 minutes -- and there would be essentially no pointing information, but the instrument's field of view would be large, with roughly a 20 degree radius. The strategy would be to search for a DM decay signal from local Galactic halo, rather than from localized targets such as galaxy clusters and the Galactic Center. A Micro-X mock observation is shown in Fig.~\ref{fig:microx}.

\subsection{The Galactic Center GeV excess}

\subsubsection{Modeling continuum gamma-ray backgrounds}

For continuum gamma-ray signals, as opposed to lines, the Galactic Center is a challenging region due to large astrophysical backgrounds; to proceed, we need a way to estimate or parameterize these backgrounds. At weak-scale energies, the dominant backgrounds come from:
\begin{itemize}
\item Cosmic ray protons striking the gas, producing neutral pions which decay to gamma rays.
\item Cosmic ray electrons upscattering starlight photons to gamma-ray energies.
\item Compact sources producing gamma-rays -- pulsars, supernova remnants, etc.
\end{itemize}
While the underlying physical processes generating these backgrounds are largely well-understood, the three-dimensional distributions of gas, starlight and cosmic rays are not well-measured, making precise prediction difficult. However, we can at least say that these backgrounds should roughly trace the distributions of gas and stars (stars can be gamma-ray point sources themselves, or generate starlight that is upscattered by cosmic-ray electrons; supernovae are thought to generate cosmic rays), which are much more dense in the disk of the Milky Way.

\begin{figure}
\centering{\hspace{10mm}
\includegraphics[width=0.9\textwidth]{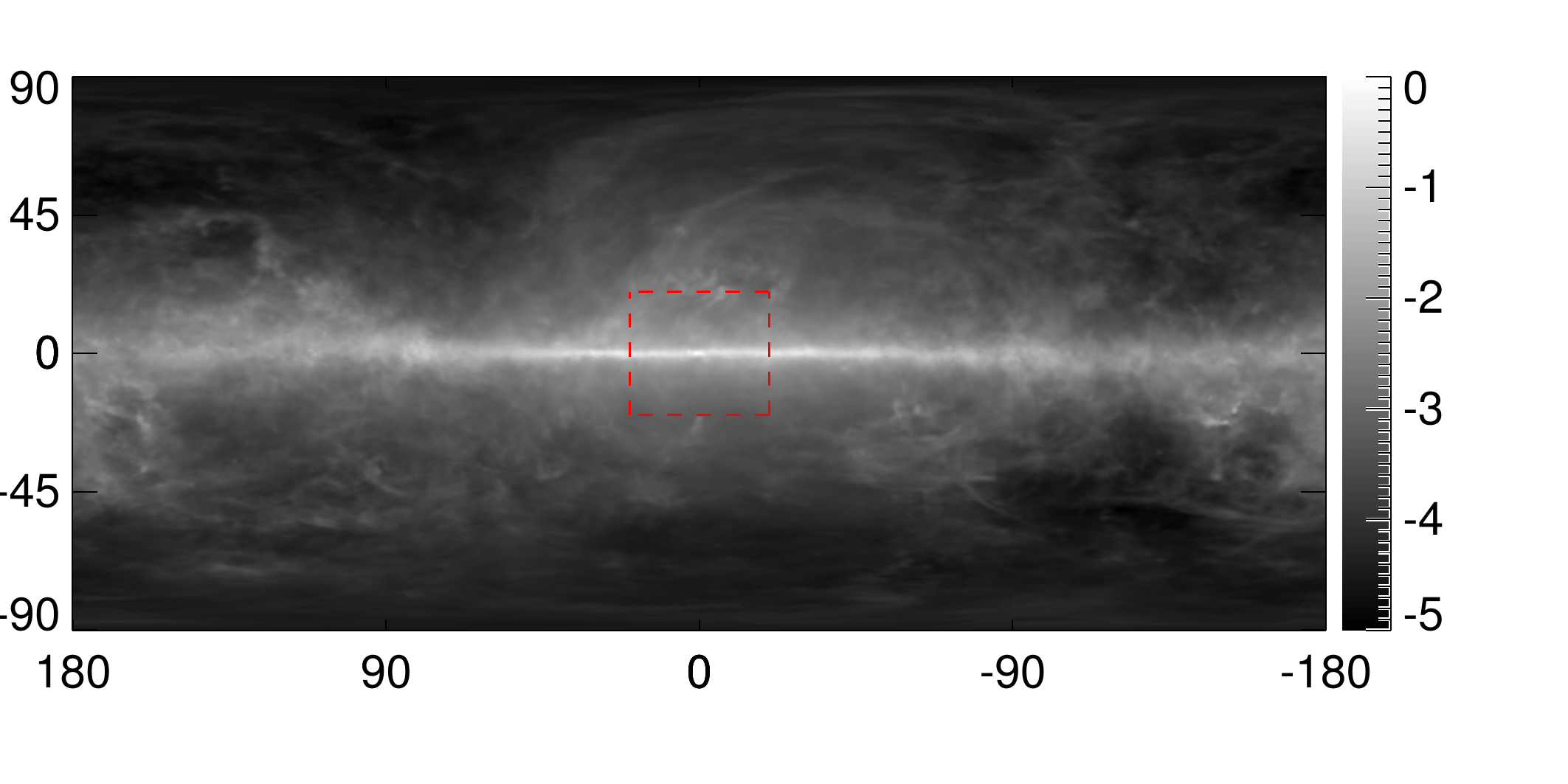}}
\\
\centering{\includegraphics[width=0.4\textwidth]{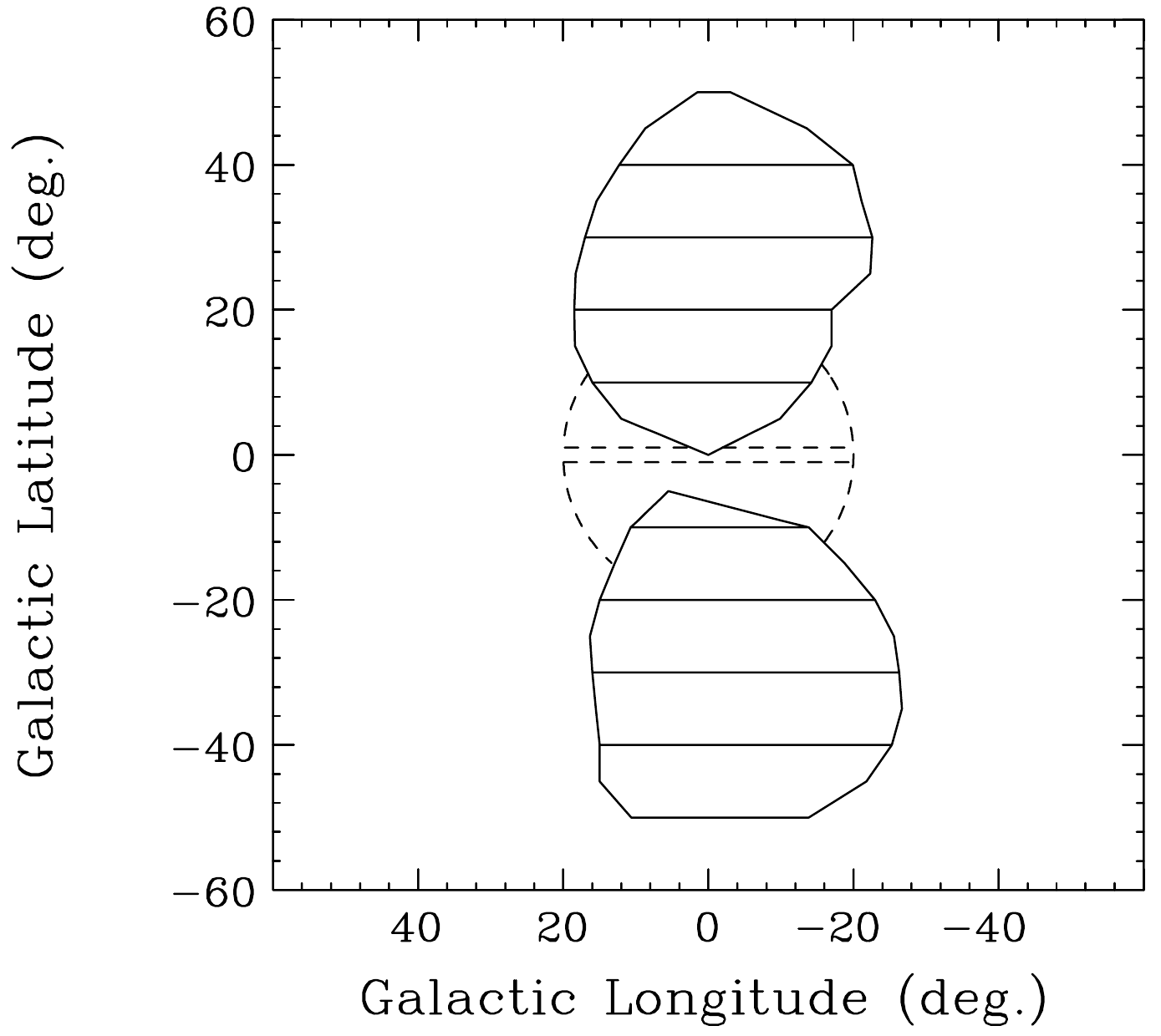} \hspace{4mm}
\includegraphics[width=0.4\textwidth]{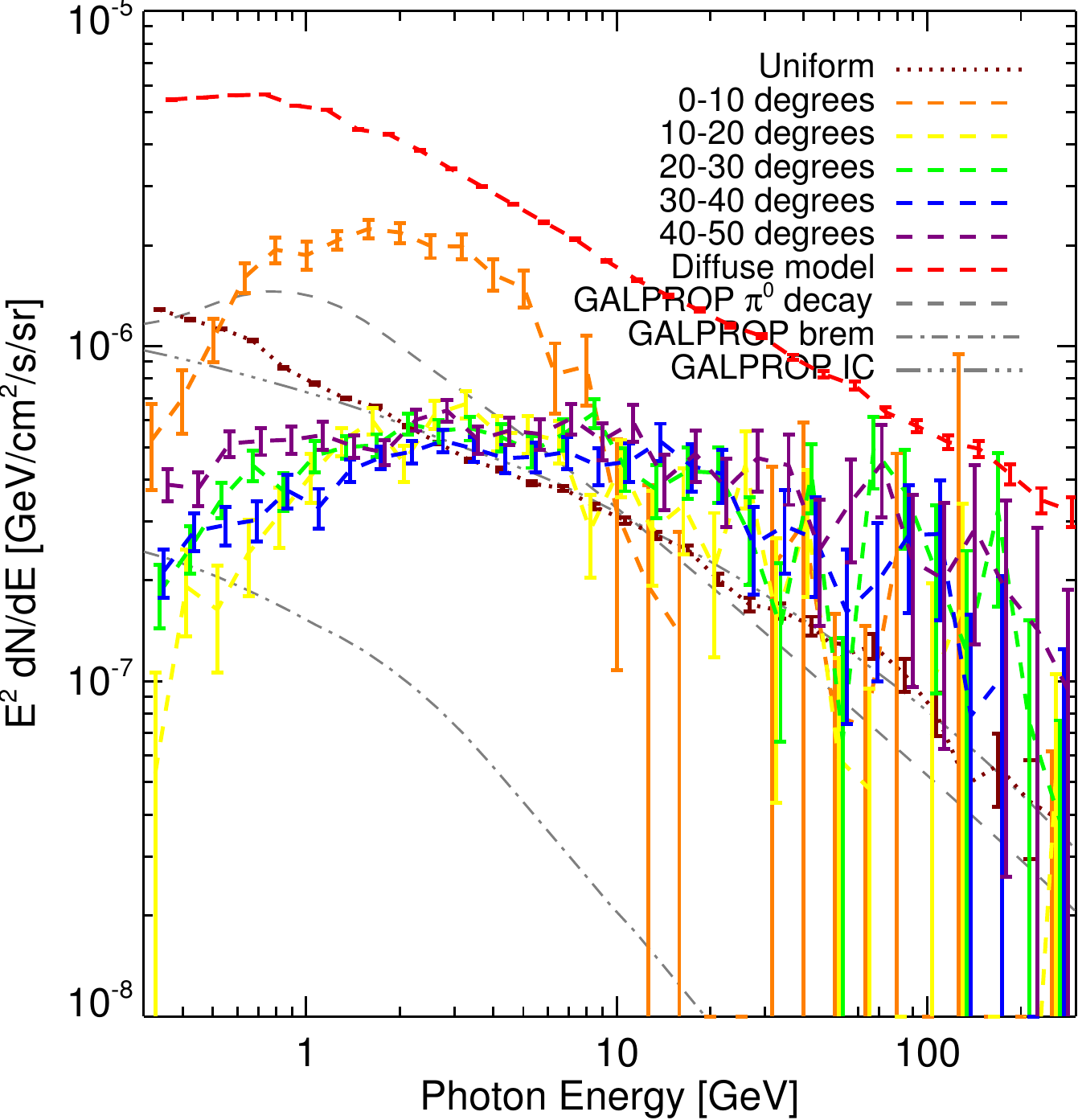}}
\caption{\emph{Top:} \texttt{p6v11} \emph{Fermi} Collaboration model for the diffuse gamma-ray emission, evaluated at an energy of 1 GeV; the red dashed lines indicate a $40\times 40^\circ$ region around the Galactic Center (reproduced from Ref.~\cite{Daylan:2014rsa}). \emph{Bottom left:} spatial template for the \emph{Fermi} Bubbles. Horizontal lines indicate ten-degree bands in Galactic latitude. Dashed lines indicate the region within $20^\circ$ of the Galactic center, but more than $1^\circ$ from the Galactic plane. \emph{Bottom right:} Spectra extracted from a template fit, modeling the sky as a linear combination of the diffuse model shown in the top panel (red dashed line), the isotropic gamma-ray background (brown dotted line), and the latitudinally sliced templates for the \emph{Fermi} Bubbles shown in the bottom left panel (orange, yellow, green, blue and purple dashed lines, for latitudes of $0-10^\circ$, $10-20^\circ$, $20-30^\circ$, $30-40^\circ$, $40-50^\circ$ respectively). Gray lines indicate the expected spectra from cosmic rays interacting with the gas and starlight. Reproduced from Ref.~\cite{Hooper:2013rwa}; see that work for details.}
\label{fig:poissonfit}
\end{figure} 

A model for the astrophysical diffuse (non-point-source) backgrounds can be constructed from maps of the gas distribution and models for the cosmic-ray and radiation distributions; for example, the latter can be taken from the public GALPROP code. There are existing public models made available by the \emph{Fermi} Collaboration \cite{FermiLAT:2012aa, Acero:2016qlg};\footnote{Current and previous models are available from the \emph{Fermi} website, \texttt{https://fermi.gsfc.nasa.gov/ssc/data/access/lat/BackgroundModels.html}} however, caution should be used when employing these models for analysis of diffuse signals, as they are designed for point source searches. The later official \emph{Fermi} models either include ad hoc spatial templates to absorb large-scale discrepancies between the initial model and the data, or re-add smoothed data-minus-model residuals (which would include any large-scale diffuse signals) to the model at the final step of processing. 

Typically one models the sky as a linear combination of spatial ``templates'', each corresponding to one or more emission mechanisms; the normalization of these templates may be fitted separately in each energy bin, or a spectrum for the template may be imposed externally and only the overall normalization fitted. This approach is not restricted to gamma-rays; similar template methods have been used in the microwave sky to remove foregrounds in studies of the CMB (e.g. Ref.~\cite{Bennett:2003ca}), and to probe possible DM signals (e.g. Refs.~\cite{Hooper:2007kb,Dobler:2008ww}). 

Fig.~\ref{fig:poissonfit} displays an example of such a fit using an early \emph{Fermi} Collaboration diffuse model (note the disk-like distribution of emission, brightest along the plane of the Milky Way) and a simple template for the large-scale gamma-ray structures known as the \emph{Fermi} Bubbles \cite{Su:2010qj}, divided into slices by latitude. The normalization of each template is allowed to float in each energy bin, allowing the extraction of a data-driven spectrum for each model component, as shown in the figure. This analysis was used in Ref.~\cite{Hooper:2013rwa} to study the spectrum of the \emph{Fermi} Bubbles as a function of latitude; the pronounced GeV-scale bump in the lowest-latitude slice is associated with the Galactic Center GeV excess, which I will discuss next.

\subsubsection{Properties of the GeV excess}

Such template-based studies indicate the presence of a new gamma-ray emission component in the Galactic Center (initially found by Refs.~\cite{Goodenough:2009gk, Hooper:2010mq}), and the inner Galaxy within $\sim 10^\circ$ of the Galactic Center (initially found by Ref.~\cite{Hooper:2013rwa}). The spectrum of this component is peaked at 1-3 GeV, and if interpreted as a possible DM signal, the size and spectrum of the signal are consistent with relatively light ($\lesssim 100$ GeV) thermal relic DM annihilating to quarks. Spatially, the signal resembles a slightly steepened NFW profile, with no flat-density core.

The morphology of this excess is highly spatially symmetric about the Galactic Center, not elongated along the plane \cite{Daylan:2014rsa,Calore:2014xka}; it also appears centered on the Galactic Center \cite{Daylan:2014rsa}. This symmetry is suggestive of an origin in the halo of the Milky Way, rather than the disk.

Recent work by the \emph{Fermi} Collaboration \cite{TheFermi-LAT:2015kwa} appears to identify the same excess. This work features a careful alternate approach to background/foreground modeling; while the spectrum of the extracted excess depends on the diffuse background modeling, the presence of a peak around a few GeV in energy is broadly robust. The greatest improvements in the fit are provided by spatial models peaked steeply toward the Galactic Center.

\subsubsection{Implications of a DM origin}

\begin{figure*}
\centerline{\includegraphics[width=0.9\textwidth]{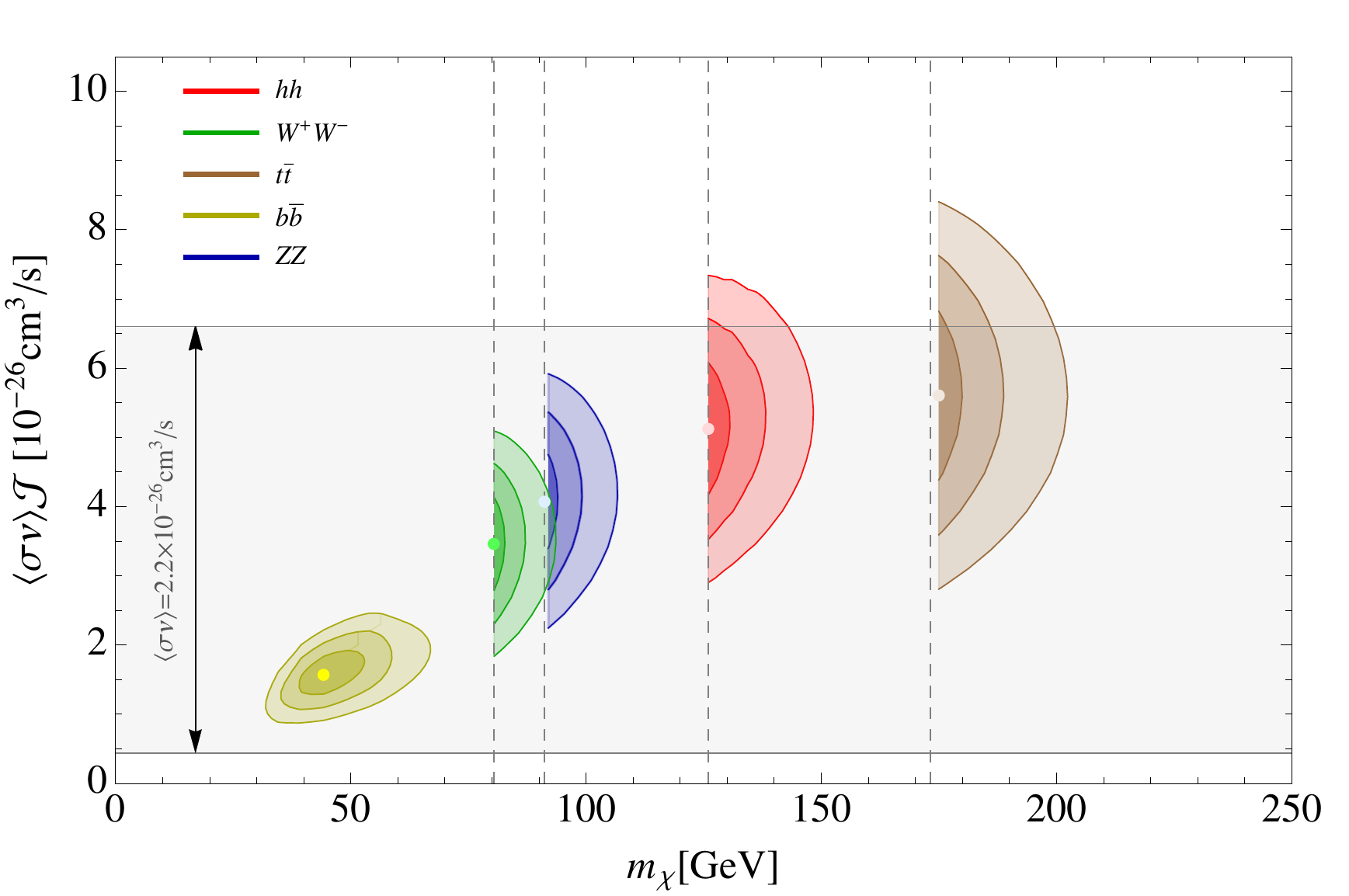}}
\caption{Preferred regions in cross section and mass for DM annihilation to various SM final states, if DM annihilation explains the full Galactic Center excess. Reproduced from Ref.~\cite{Agrawal:2014oha}.}
\label{fig:gcewimps}
\end{figure*}

If the excess originates from DM, the greatest improvement in the fit occurs for DM masses around 10-100 GeV depending on the annihilation channel. For $b$ quarks, the overall best-fit channel, the best-fit mass is $\sim 40-50$ GeV. The required cross section is close to thermal, i.e. approximately weak-scale. Heavier DM annihilating to $\bar{h} h$ can also provide a good fit \cite{Agrawal:2014oha}, but the preferred DM mass is right at the threshold for $\bar{h} h$ production. Annihilation to W bosons, Z bosons and top quarks provides a slightly worse fit; again the preferred mass is close to threshold, as shown in Fig.~\ref{fig:gcewimps}.

There are non-negligible model-building challenges for a DM explanation of this signal, although there are existence proofs of UV-complete models that satisfy all constraints. Direct detection is very sensitive in this mass range, so the lack of a detection must be explained. Some possibilities include a resonant enhancement to the annihilation rate, a suppression of the direct detection rate due to spin-dependence or some other effect (although upcoming direct-detection experiments may have sensitivity anyway), or a scenario where the annihilation is to intermediate particles that subsequently decay into visible particles with a small coupling. In this third case, the small coupling between the dark sector and SM suppresses the direct detection rate, but not the annihilation rate.

There are also important limits from colliders, ruling out substantial classes of simplified models \cite{Berlin:2014tja}. The sensitivity of collider searches is reduced in the presence of light mediators, which may be needed to raise the cross section to thermal relic values.

Two example classes of viable models are:
\begin{itemize}
\item Annihilation through a pseudoscalar to $b$ quarks, e.g. the ``coy DM'' of Ref.~\cite{Boehm:2014hva}.  A renormalizable model, where the pseudoscalar mixes with the CP-odd component of a two-Higgs-doublet model, was presented in Ref.~\cite{Ipek:2014gua}. An implementation in the $Z_3$ NMSSM was worked out in Ref.~\cite{Cheung:2014lqa}, where bino/higgsino DM annihilates through a light MSSM-like pseudoscalar. A general NMSSM study was performed in Ref.~\cite{Cahill-Rowley:2014ora}.
\item $2\rightarrow 4$ models, where the DM annihilates to an on-shell mediator, which subsequently decays to SM particles, e.g. Refs.~\cite{Ko:2014gha, Abdullah:2014lla, Martin:2014sxa}. Dark-photon and NMSSM implementations are discussed in Ref.~\cite{Berlin:2014pya}; an extension with dark-sector showering is presented in Ref.~\cite{Freytsis:2014sua}.
\end{itemize}

\subsubsection{Non-DM possibilities}

\begin{figure}
\centerline{\includegraphics[width=0.5\textwidth]{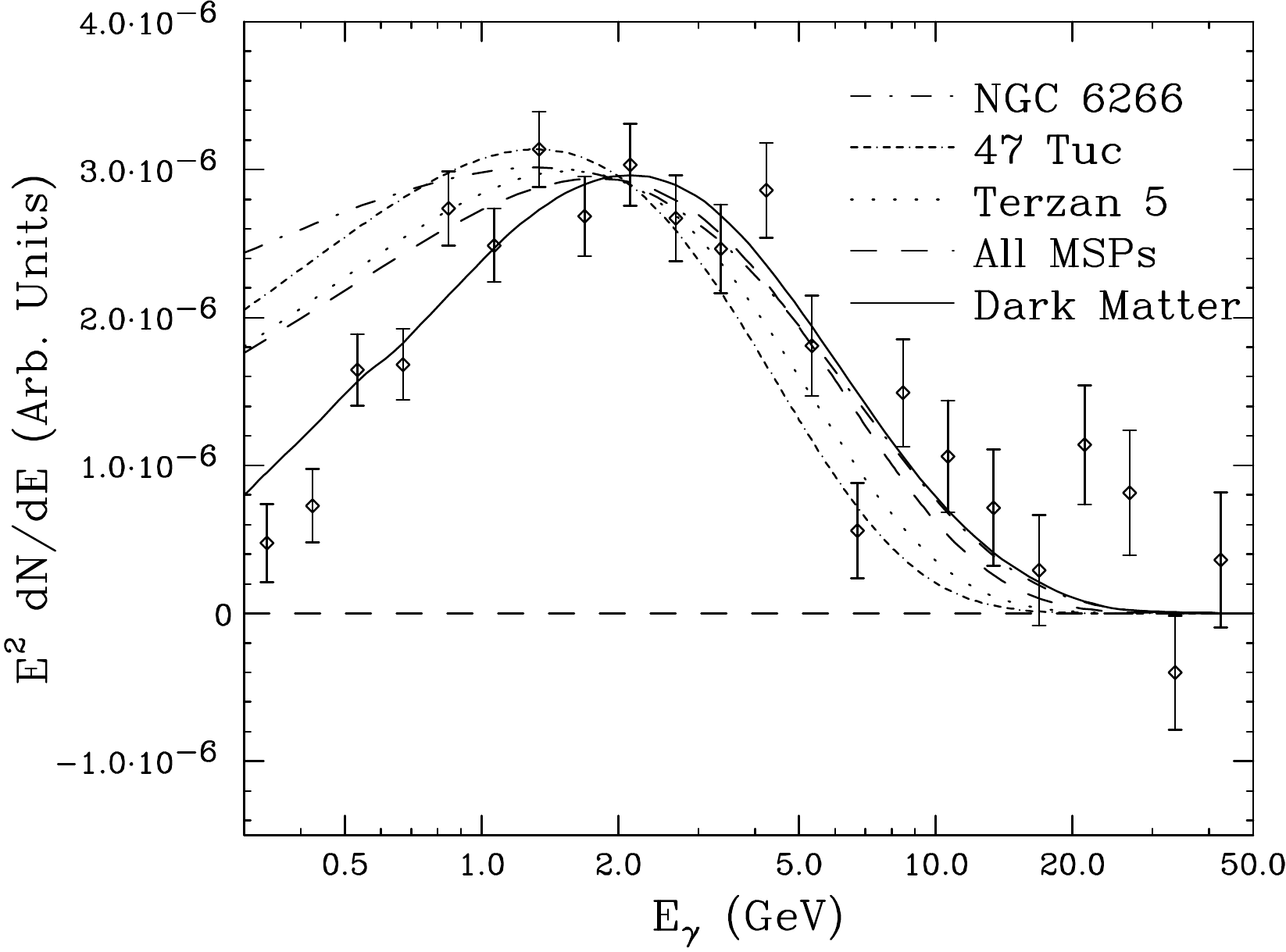}}
\caption{Observed spectrum of the GeV excess in the analysis of Ref.~\cite{Daylan:2014rsa}, compared to spectra of observed millisecond pulsars and several globular clusters (whose gamma-ray emission is thought to be dominated by millisecond pulsars. Reproduced from Ref.~\cite{Daylan:2014rsa}; see that work for details.}
\label{fig:pulsars}
\end{figure}

Pulsars, spinning neutron stars, are known to emit gamma rays with a very similar spectrum to the observed excess, as shown in Fig.~\ref{fig:pulsars}. There is no strong a priori reason to expect pulsars to have a spatial distribution matching the excess, but of course this does not rule it out; at least one recent study has suggested a possible mechanism for generating a spherical pulsar population peaked toward the Galactic Center \cite{Brandt:2015ula}.

Outflows of high-energy cosmic rays from the Galactic Center could also produce gamma rays, as discussed previously. However, it would be surprising if the excess originated wholly from protons interacting with the gas, as the signal does not appear to be gas-correlated. Electrons upscattering photons to gamma-ray energies could also contribute, although multiple sources may be required to accommodate an electron spectrum that does not change markedly with position. For discussion of these points and more, see e.g. Refs.~\cite{Carlson:2014cwa, Petrovic:2014uda, Cholis:2015dea, Gaggero:2015nsa}.

\subsubsection{Photon statistics}

One way to distinguish between these various hypotheses is to examine the \emph{clumpiness} of the photons. If the signal originates from DM annihilation or an outflow, we would expect to observe a fairly smooth spatial distribution of flux. In the pulsar case, we might instead see many ``hot spots'' scattered over a fainter background. This general claim can be made quantitative by considering the differing photon statistics in these two cases; the variance is larger for a given mean when point sources are present, and this can be captured in a modification to the likelihood function, even if the precise locations of the point sources are not known. The method I discuss here is described in more detail in Ref.~\cite{Lee:2015fea}; a related analysis using wavelet techniques \cite{Bartels:2015aea} finds consistent results.

As a simple example of the required modification to the likelihood in the presence of point sources, consider a scenario where 10 photons per pixel are expected, in some region of the sky. What is the probability of finding 0 photons? 12 photons? 100 photons?

In a case where only diffuse emission is present, Poissonian statistics are valid; the probability of observing exactly 12 photons is $P(12) = 10^{12} e^{-10}/12! \sim 0.1$, and likewise $P(0)\sim 5\times 10^{-5}$, $P(100) \sim 5\times 10^{-63}$. But suppose instead we are told that the source of emission is a population of rare sources, each of which produces 100 photons on average, but with only 0.1 sources expected per pixel. Poissonian statistics now cannot be applied, as observing a photon from a given pixel tells us that there is a source present there, and increases the probability of seeing another photon from the same pixel -- the events are no longer independent.  The mean number of photons per pixel remains the same, but now the probability of seeing zero photons is $P(0) \sim 0.9$ (neglecting the possibility that a source is present but fluctuates down to 0 photons from the 100 expected), $P(12) \sim 0.1\times100^{12} e^{-100}/12! \sim 10^{-29}$, and $P(100) \sim 4\times 10^{-3}$ (in the latter two cases, I have neglected terms arising from the case where multiple sources are present in a pixel, as this is rare). 

Thus the expected \emph{distribution} of the number of photons is very different, even though the mean is the same, between the two cases. In the first case, seeing 12 photons is quite likely, but in the second case it is essentially impossible, since this would require a source to be present but to produce only 12 photons when 100 are expected. Likewise, observing 100 photons will never happen in the first case, but is quite plausible in the second, if there are a few hundred pixels with these properties. 

\begin{figure*}[t]
	\centering{\includegraphics[width=1.0\textwidth]{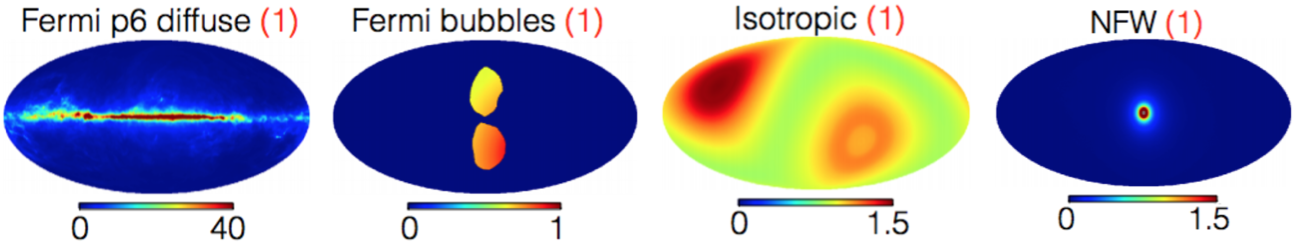} \\
	\includegraphics[width=0.75\textwidth]{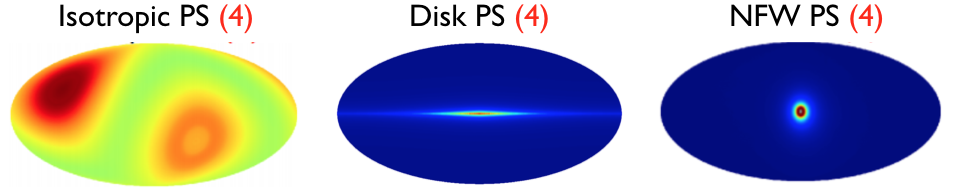}} 
	\caption{Skymaps for templates for gamma-ray diffuse emission (top row) and gamma-ray point sources (bottom row). See text for details.}
	\label{fig:nptemplates}
\end{figure*}

In the template fitting method discussed earlier, each template was assumed to possess Poissonian statistics. We can now extend this method to \emph{non-Poissonian} template fitting, and thus include templates corresponding to populations of (potentially unresolved) sources. The overall spatial distribution of the sources is specified as previously, but now the probability of observing a certain number of photons, given a model for the total number of photons, is determined via non-Poissonian statistics (appropriate to a point source population, as above) rather than Poissonian. Fig.~\ref{fig:nptemplates} displays the templates we will use for diffuse and point-source emission components. We model the point-source population as a combination of isotropic extragalactic sources, Galactic sources tracing the disk of the Milky Way, and (optionally) point sources following the spatial distribution of the GeV excess; the diffuse emission is modeled as a combination of the \emph{Fermi} \texttt{p6v11} diffuse model, the \emph{Fermi} Bubbles, the diffuse isotropic gamma-ray background, and a DM-like template following the spatial distribution of the GeV excess.

As in the example above, the non-Poissonian probability of observing a certain number of photons also depends on the properties of the source population: specifically, the number of sources as a function of their brightness, the \emph{source count function}. For each non-Poissonian template, we can add extra parameters to describe the source count function, and fit for these parameters just as we fit for the normalizations of the templates. As a default, we treat the source count function as a broken power law described by three parameters; the indices above and below the break, and the flux at which the break occurs.

\begin{figure*}[t]
	\centerline{\includegraphics[width=0.5\textwidth]{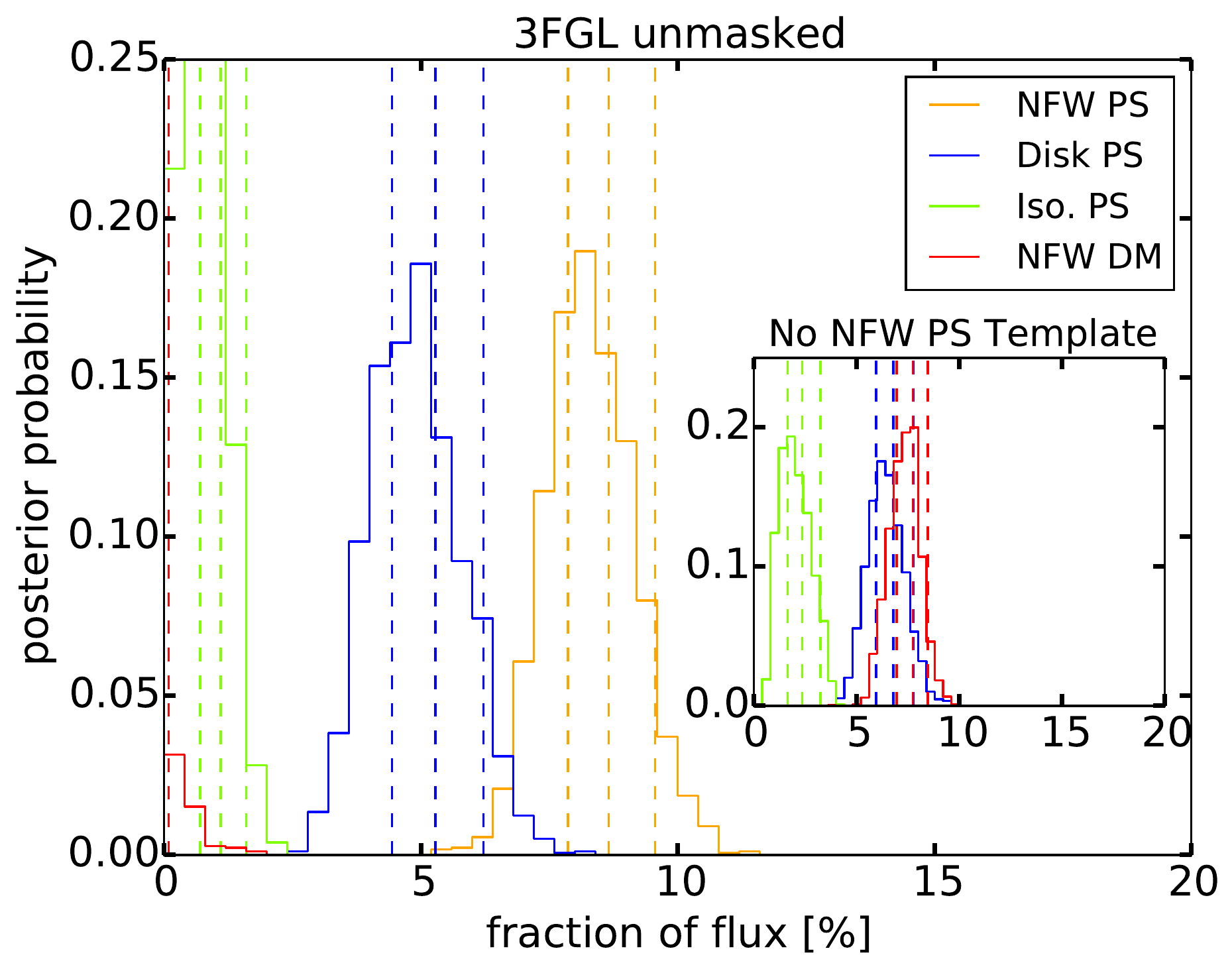}} 
	\caption{Posteriors for the flux fraction within $10^\circ$ of the Galactic Center with $|b| \geq 2^\circ$ arising from the separate point source components, with known point sources unmasked.  The inset shows the result of removing the NFW PS template from the fit.  Dashed vertical lines indicate the $16^\text{th}$, $50^\text{th}$, and $84^\text{th}$ percentiles. Reproduced from Ref.~\cite{Lee:2015fea}.}
	\label{fig:npresults}
\end{figure*}

We can then perform fits of two models, one including a template for point sources tracing the GeV excess (labeled ``NFW PS''), and one without this template. In both cases we include the smooth ``NFW DM'' template with morphology characteristic of the GeV excess. We found\cite{Lee:2015fea} that if the ``NFW PS'' template is absent, the ``NFW DM'' template absorbs the GeV excess, as previously. But when the ``NFW PS'' template is added to the fit, then it absorbs the full excess, driving the ``NFW DM'' template to zero, as shown in Fig.~\ref{fig:npresults}. 

This result suggests that the GeV excess can be better explained by a population of point sources than by smooth, diffuse emission, whether from DM or from an outflow of cosmic rays.

The leading candidate for these point sources is a population of pulsars, as briefly discussed above, given their spectral similarity to the GeV excess. Such a population could potentially be detected at other frequencies, e.g. by radio or X-ray telescopes; in particular, prospects for detection with the MeerKAT or SKA telescopes have been studied in Ref.~\cite{Calore:2015bsx}.

\section*{Acknowledgements}
These lectures were originally presented at TASI 2016: Anticipating the Next Discoveries in Particle Physics, which was supported by the U.S.~National Science Foundation under Grants No.~1305809 and No.~1505221. I thank the organizers of TASI 2016 -- Rouven Essig, Ian Low,  and Tom DeGrand -- for their hard work and for giving me the opportunity to lecture, and all the students of TASI 2016 for their invaluable questions and ideas. My work is conducted with support from the U.S. Department of Energy under grant Contract Numbers DE-SC00012567 and DE-SC0013999.

\bibliography{tasi-bib}
\end{document}